\documentclass[fleqn,usenatbib]{mnras} %
\usepackage[multidot]{grffile}

\usepackage[utf8]{inputenc}
\usepackage[T1]{fontenc}
\usepackage{ae,aecompl}

\usepackage{graphicx}	
\usepackage{amsmath}	
\usepackage{amssymb}	

\usepackage{wasysym}
\usepackage{courier}
\usepackage{mathrsfs}

\usepackage{threeparttable}
\usepackage{booktabs}
\usepackage{longtable}

\usepackage{verbatim}
\usepackage{comment}
\usepackage[british]{babel}
\usepackage{color}
\usepackage{booktabs}
\usepackage{tabularx}

\newcommand{\beq}{\begin{equation}}
\newcommand{\eeq}{\end{equation}}
\newcommand{\kepler}[0]{\emph{Kepler}}
\newcommand{\corot}[0]{\emph{CoRoT}}
\newcommand{\teff}[0]{$T_\text{eff}$}
\newcommand{\logg}[0]{$\log g$}
\newcommand{\Dnu}[0]{$\Delta\nu$}
\newcommand{\dnu}[1]{$\delta\nu_{#1}$}
\newcommand{\numax}[0]{$\nu_{\rm max}$}
\newcommand{\dpi}[1]{$\Delta\Pi_{#1}$}
\newcommand{\amax}[0]{$A_{\rm max}$}
\newcommand{\epsp}[0]{$\epsilon_p$}
\newcommand{\epsg}[0]{$\epsilon_g$}
\newcommand{\nug}[1]{$\gamma_{#1}$}

\newcommand{\muhz}[0]{$\mu{\rm Hz}$}
\newcommand{\tobs}[0]{$t_{\rm obs}$}

\newcommand{\lwmax}[0]{$\Gamma(\nu_{\rm max})$}
\newcommand{\echelle}[0]{\'{e}chelle}

\newcommand{\refa}[0]{\citet{appourchaux++2012-freq-ms-sg-kepler}} 
\newcommand{\refb}[0]{\citet{deheuvels++2014-internal-rotation-6-sg-rg-kepler}} 
\newcommand{\refc}[0]{\citet{tianzj-2015-subgiant-kic6442183-kic11137075}} 
\newcommand{\refd}[0]{\citet{appourchaux++2014-lw-height-ms-kepler}} 
\newcommand{\refe}[0]{\citet{davies++2016-bayesian-peakbagging-35-kepler-planethost}} 
\newcommand{\reff}[0]{\citet{campante++2011-subgiants-kic10273246-kic10920273}} 
\newcommand{\refg}[0]{\citet{mathur++2011-kic1395018-kic11234888-solartype}}

\title[Asteroseismology of 36 {\em Kepler} subgiants -- I]{Asteroseismology of 36 \kepler{} subgiants -- I. Oscillation frequencies, linewidths and amplitudes}

\author[Li et al.]{%
Yaguang Li,$^{1,2,3}$\thanks{yaguang.li@mail.bnu.edu.cn}
Timothy R. Bedding,$^{2,3}$ \thanks{tim.bedding@sydney.edu.au}
Tanda Li,$^{2,3}$ 
Shaolan Bi$^{1}$ \thanks{bisl@bnu.edu.cn}
\newauthor
Dennis Stello,$^{4,2,3}$
Yixiao Zhou$^{5}$ and
Timothy R. White$^{2,3}$
\\
$^{1}$Department of Astronomy, Beijing Normal University, Beijng 100875, China\\
$^{2}$Sydney Institute for Astronomy (SIfA), School of Physics, University of Sydney, NSW 2006, Australia\\
$^{3}$Stellar Astrophysics Centre, Department of Physics and Astronomy, Aarhus University, Ny Munkegade 120, \\
\ DK-8000 Aarhus C, Denmark\\
$^{4}$School of Physics, University of New South Wales, 2052, Australia\\
$^{5}$Research School of Astronomy and Astrophysics, Australian National University, Canberra, ACT 2611, Australia\\
}

\date{Accepted in xxx. Received in yyy; in original form zzz}

\pubyear{2018}

\begin{document}
\label{firstpage}
\maketitle

\begin{abstract}
The presence of mixed modes makes subgiants excellent targets for asteroseismology, providing a probe for the internal structure of stars. Here we study 36 \emph{Kepler} subgiants with solar-like oscillations and report their oscillation mode parameters. We performed a so-called peakbagging exercise, i.e. estimating oscillation mode frequencies, linewidths, and amplitudes with a power spectrum model, fitted in the Bayesian framework and sampled with a Markov Chain Monte Carlo algorithm. The uncertainties of the mode frequencies have a median value of 0.180 $\mu$Hz. We obtained seismic parameters from the peakbagging, analysed their correlation with stellar parameters, and examined against scaling relations. The behaviour of seismic parameters (e.g. $\Delta\nu$, $\nu_{\rm max}$, $\epsilon_p$) is in general consistent with theoretical predictions. We presented the observational p--g diagrams: $\gamma_1$--$\Delta\nu$ for early subgiants and $\Delta\Pi_1$--$\Delta\nu$ for late subgiants, and demonstrate their capability to estimate stellar mass. We also found a $\log g$ dependence on the linewidths and a mass dependence on the oscillation amplitudes and the widths of oscillation excess. This sample will be valuable constraints for modelling stars and studying mode physics such as excitation and damping.  
\end{abstract}



\section{Introduction}
\label{sec:intro}

Some of the first detections of solar-like oscillations, using ground-based spectroscopy, were in subgiant stars such as $\eta$ Boo \citep{kjeldsen++1995-eta-bootis,jcd++1995-eta-boo,guenther+1996-eta-boo,kjeldsen++2003-eta-boo,dimauro++2003-eta-boo,guenther-2004-eta-boo,carrier++2005-eta-boo}, $\beta$ Hyi \citep{bedding++2001-betaHydri-evidence,carrier++2001-beta-hyi,dimauro++2003-beta-hyi,fernandes+2003-beta-hyi,bedding++2007-beta-hyi,brandao++2011-beta-hyi}, $\nu$ Ind \citep{bedding++2006-nu-ind,carrier++2007-nu-ind} and $\mu$ Her \citep{bonanno++2008-mu-her,pinheiro+2010-mu-her,grundahl++2017-mu-her}. Observations by the space telescopes \corot{} and \kepler{} enabled high-quality photometry with high duty cycle. These include subgiants such as HD 49385 \citep{deheuvels+2010-hd49385-corot-interior-model,paxton++2013-mesa-planets-oscillations-rotation-massivestars}, HD 169392A \citep{mathur++2013-hd169392a}, KIC 11026764 \citep[``Gemma'';][]{metcalfe++2010-kic11026764-sg}, KIC 11395018 \citep[``Boogie'';][]{mathur++2011-kic1395018-kic11234888-solartype}, KIC 10920273 and KIC 10273246 \citep[``Scully'' and ``Mulder'';][]{campante++2011-subgiants-kic10273246-kic10920273}, the $\alpha$-enhanced star KIC 7976303 \citep{ge++2015-two-alpha-enhanced-stars-kic7976303-kic8694723} and the binary twin system KIC 7107778 \citep{liyg++2018-kic7107778}.

Beginning with $\eta$ Boo, it was realised that subgiants show mixed modes \citep{jcd++1995-eta-boo}, which are now recognised as a feature in all evolved stars. Main-sequence (MS) dwarfs with solar-like oscillations host a convective envelope, which excites and propagates pressure (p) modes. In subgiants, the so-called mixed modes that result from coupling between the pressure mode cavity and gravity (g) mode cavity carry information from the core and have observable amplitudes on the surface.

The mixed modes have long been realised to have strong diagnostic potential because the frequencies of g modes, which we denote by \nug{}, evolve quite rapidly \citep{jcd++1995-eta-boo}. \citet{bedding-2014-solar-like-review} suggested a new asteroseismic diagram, the p\,--\,g diagram, which plots \nug{} versus the large separation of p modes \Dnu{}. By comparing observed values of \nug{} and \Dnu{} with stellar models on this diagram, a stellar mass can be estimated. \citet{benomar++2012-mass-sg-kepler} showed that the coupling strength (see Section~\ref{sec:parameter} below) is a strong indicator of evolutionary stage in subgiants. These suggestions of the diagnostic power of mixed modes were confirmed with modelling. When modelling Gemma (KIC 11026764), \citet{metcalfe++2010-kic11026764-sg} showed that the asteroseismic age could be constrained to a precision of 15\% that was set by the choice of input physics. \citet{litd++2019-mu-her-song} modelled $\mu$ Her and demonstrated that the asteroseismic age did not significantly change as a function of the mixing-length parameter, or the initial helium abundance. In addition to global parameters, the mixed modes also shed light on the interior physics of stars, for example, constraining the buoyancy frequency profile \citep{litd++2019-mu-her-song}, internal angular momentum transport \citep{deheuvels++2014-internal-rotation-6-sg-rg-kepler,eggenberger++2018-am-transport-sg} and convective core overshooting \citep{deheuvels++2016-overshoot}.

A complete analysis of oscillation modes has been made for \kepler{} observations of 35 planet-host stars \citep{davies++2016-bayesian-peakbagging-35-kepler-planethost} and 66 MS stars \citep{lund++2017-legacy-kepler-1}, the so-called LEGACY sample. However, similar analyses for subgiants has only been done on a subset of \kepler{} data \citep[e.g.][]{appourchaux++2012-freq-ms-sg-kepler}. Therefore it is valuable to gather and analyse subgiants with the full set of \kepler{} observations, given the great promise of mixed modes. In this paper, we fit the oscillation modes and analyse the fitted frequencies, linewidths and amplitudes from the fit. In a companion paper (Paper II, T. Li et al. submitted), we perform a detailed modelling to these stars with the constraints of frequencies obtained here.

\section{Observations}
\label{sec:target}

\begin{table*} 
\caption{Observational properties.} 
\label{tab:atm} 
\begin{tabular*}{\textwidth}{@{\extracolsep{\fill}}rlrlrrl}
\toprule
      KIC &  Nickname & $t_\text{obs}$ (d) &  $T_\text{eff}$ (K) & $\log{g}$ (cgs; dex) &          [Fe/H] (dex) & References \\
\midrule
  2991448 &      ---  &              $204$ &   $5623$ $\pm$ $80$ &  $3.98$ $\pm$ $0.02$ &  $-0.10$ $\pm$ $0.15$ &       ---  \\
  3852594 &      ---  &              $540$ &   $6296$ $\pm$ $78$ &  $4.02$ $\pm$ $0.14$ &  $-0.40$ $\pm$ $0.15$ &       ---  \\
  4346201 &      ---  &              $108$ &   $6058$ $\pm$ $81$ &  $3.97$ $\pm$ $0.01$ &  $-0.24$ $\pm$ $0.17$ &       ---  \\
  5108214 &      ---  &              $222$ &   $5799$ $\pm$ $78$ &  $3.81$ $\pm$ $0.07$ &   $0.16$ $\pm$ $0.15$ &       ---  \\
  5607242 &      ---  &             $1031$ &   $5485$ $\pm$ $81$ &  $3.76$ $\pm$ $0.06$ &  $-0.06$ $\pm$ $0.15$ &          a \\
  5689820 &      ---  &              $506$ &   $5037$ $\pm$ $76$ &  $3.76$ $\pm$ $0.06$ &   $0.21$ $\pm$ $0.15$ &          b \\
  5955122 &  Rhapsody &              $988$ &   $5877$ $\pm$ $79$ &  $3.87$ $\pm$ $0.01$ &  $-0.22$ $\pm$ $0.15$ &          a \\
  6064910 &      ---  &              $532$ &   $6376$ $\pm$ $80$ &  $3.83$ $\pm$ $0.01$ &  $-0.26$ $\pm$ $0.17$ &       ---  \\
  6370489 &      ---  &               $97$ &   $6184$ $\pm$ $80$ &  $3.92$ $\pm$ $0.01$ &  $-0.36$ $\pm$ $0.15$ &       ---  \\
  6442183 &    Dougal &              $822$ &   $5702$ $\pm$ $76$ &  $4.00$ $\pm$ $0.01$ &  $-0.20$ $\pm$ $0.15$ &       c, d \\
  6693861 &      ---  &              $565$ &   $5626$ $\pm$ $84$ &  $3.84$ $\pm$ $0.02$ &  $-0.36$ $\pm$ $0.15$ &       ---  \\
  6766513 &      ---  &              $534$ &   $6227$ $\pm$ $78$ &  $3.93$ $\pm$ $0.02$ &  $-0.18$ $\pm$ $0.15$ &       ---  \\
  7174707 &      ---  &             $1031$ &   $5168$ $\pm$ $82$ &  $3.71$ $\pm$ $0.12$ &   $0.07$ $\pm$ $0.15$ &       ---  \\
  7199397 &      ---  &             $1182$ &   $5903$ $\pm$ $79$ &  $3.76$ $\pm$ $0.07$ &  $-0.14$ $\pm$ $0.15$ &          e \\
  7668623 &      ---  &              $112$ &   $6228$ $\pm$ $77$ &  $3.87$ $\pm$ $0.02$ &   $0.24$ $\pm$ $0.15$ &       ---  \\
  7747078 &      ---  &              $986$ &   $5903$ $\pm$ $74$ &  $3.90$ $\pm$ $0.01$ &  $-0.22$ $\pm$ $0.15$ &          a \\
  7976303 &      John &              $962$ &   $6079$ $\pm$ $81$ &  $3.89$ $\pm$ $0.06$ &  $-0.48$ $\pm$ $0.15$ &          a \\
  8026226 &     Gypsy &              $305$ &   $6224$ $\pm$ $84$ &  $3.70$ $\pm$ $0.06$ &  $-0.18$ $\pm$ $0.17$ &          a \\
  8524425 &   Katrina &             $1006$ &   $5543$ $\pm$ $87$ &  $3.97$ $\pm$ $0.01$ &   $0.02$ $\pm$ $0.17$ &          a \\
  8702606 &      ---  &              $980$ &   $5529$ $\pm$ $82$ &  $3.76$ $\pm$ $0.06$ &  $-0.16$ $\pm$ $0.15$ &       a, b \\
  8738809 &      ---  &              $341$ &   $6045$ $\pm$ $72$ &  $3.90$ $\pm$ $0.01$ &   $0.20$ $\pm$ $0.12$ &       ---  \\
  9512063 &      ---  &              $110$ &   $5838$ $\pm$ $78$ &  $3.88$ $\pm$ $0.02$ &  $-0.22$ $\pm$ $0.15$ &       ---  \\
 10018963 &     Klaas &              $936$ &   $6177$ $\pm$ $82$ &  $3.93$ $\pm$ $0.01$ &  $-0.22$ $\pm$ $0.17$ &          a \\
 10147635 &      ---  &              $761$ &   $5941$ $\pm$ $80$ &  $3.75$ $\pm$ $0.02$ &  $-0.02$ $\pm$ $0.15$ &       ---  \\
 10273246 &    Mulder &              $705$ &  $6269$ $\pm$ $124$ &  $4.41$ $\pm$ $0.07$ &   $0.21$ $\pm$ $0.15$ &          f \\
 10593351 &      ---  &              $528$ &   $5754$ $\pm$ $82$ &  $3.66$ $\pm$ $0.06$ &   $0.16$ $\pm$ $0.15$ &       ---  \\
 10873176 &      ---  &              $122$ &   $6520$ $\pm$ $87$ &  $3.90$ $\pm$ $0.01$ &  $-0.20$ $\pm$ $0.28$ &       ---  \\
 10920273 &    Scully &              $532$ &   $5365$ $\pm$ $85$ &  $3.78$ $\pm$ $0.12$ &  $-0.16$ $\pm$ $0.15$ &          f \\
 10972873 &      ---  &              $878$ &   $5705$ $\pm$ $81$ &  $3.96$ $\pm$ $0.02$ &  $-0.08$ $\pm$ $0.15$ &       ---  \\
 11026764 &     Gemma &              $985$ &   $5636$ $\pm$ $80$ &  $3.89$ $\pm$ $0.06$ &   $0.04$ $\pm$ $0.15$ &          a \\
 11137075 &   Zebedee &              $530$ &   $5510$ $\pm$ $74$ &  $4.00$ $\pm$ $0.01$ &  $-0.12$ $\pm$ $0.12$ &          c \\
 11193681 &      ---  &             $1026$ &   $5575$ $\pm$ $79$ &  $3.79$ $\pm$ $0.06$ &   $0.21$ $\pm$ $0.15$ &          a \\
 11395018 &    Boogie &              $562$ &  $5753$ $\pm$ $114$ &  $3.65$ $\pm$ $0.18$ &   $0.02$ $\pm$ $0.15$ &       a, g \\
 11414712 &    Jingle &              $943$ &   $5622$ $\pm$ $80$ &  $3.80$ $\pm$ $0.06$ &  $-0.14$ $\pm$ $0.15$ &          a \\
 11771760 &      ---  &             $1030$ &   $5796$ $\pm$ $78$ &  $3.67$ $\pm$ $0.06$ &  $-0.06$ $\pm$ $0.17$ &          a \\
 12508433 &      ---  &              $999$ &   $5303$ $\pm$ $78$ &  $3.83$ $\pm$ $0.06$ &   $0.20$ $\pm$ $0.15$ &       a, b \\
\bottomrule
\end{tabular*}
 \begin{tablenotes}  
 \item \emph{Note}: (a) \refa{}; (b) \refb{}; (c) \refc{}; (d) \refd{}; (e) \refe{}; (f) \reff{}; (g) \refg{}.
 \end{tablenotes}
\end{table*} 

\begin{figure}
	\includegraphics[width=\linewidth]{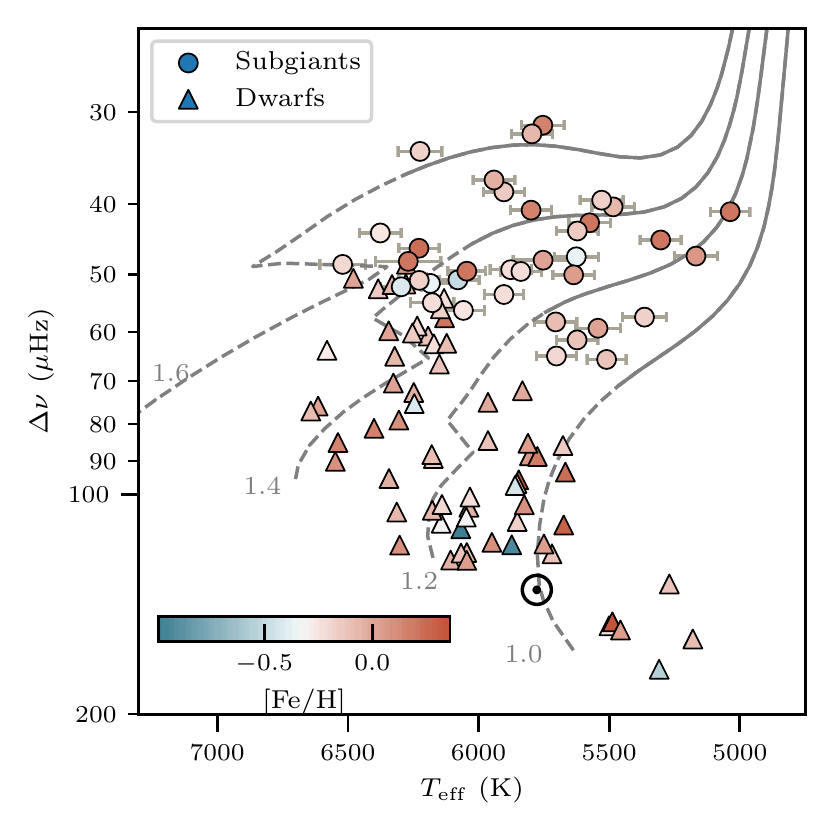}
    \caption{Asteroseismic H-R diagram showing \Dnu{} vs. \teff{} for the 36 subgiants shown as black circles colour-coded by metallicity. For comparison, the LEGACY sample \citep{lund++2017-legacy-kepler-1} is together shown in triangles labelled as dwarfs. The Sun is marked by the usual symbol. The theoretical evolutionary tracks with solar metallicity \citep{stello-2013-classifcation-13000-rg-kepler} are shown before (dashed lines) and after (solid lines) the exhaustion of central hydrogen, with each track labelled with mass in solar units.}
    \label{fig:hrd}
\end{figure}

\begin{figure}
	\includegraphics[width=\linewidth]{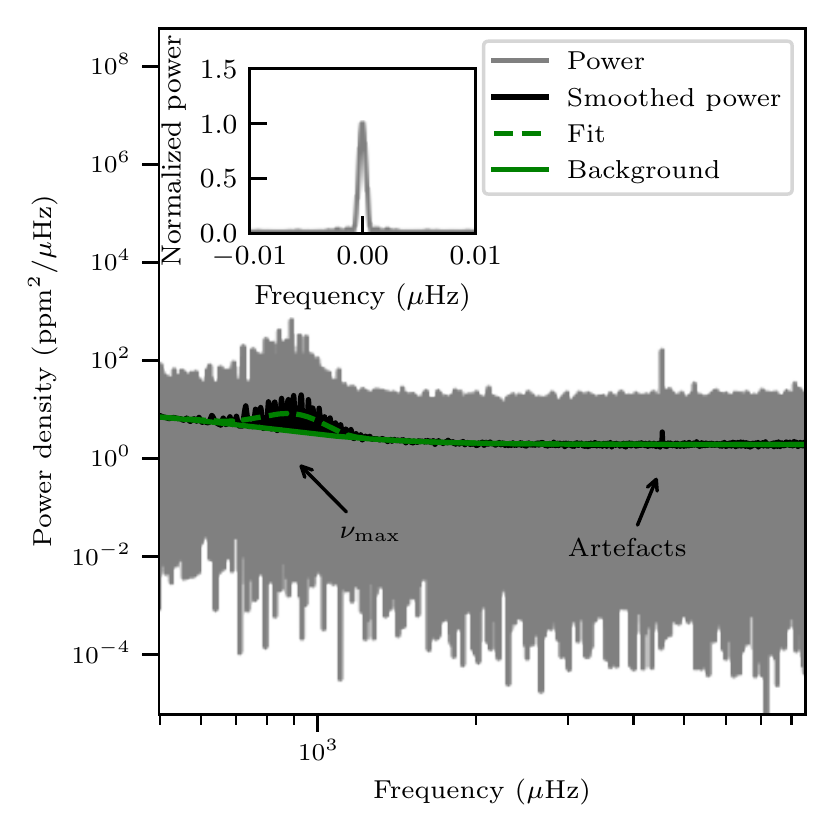}
    \caption{Power spectrum density of Gemma (KIC 11026764). The original (grey), smoothed (black) and fitted power spectra (green) are shown. The peaks around 4500 \muhz{} are artifacts \citep{Gilliland++2010-kepler-sc-data-feature}. The inset shows the normalised spectral window.}
    \label{fig:gemma}
\end{figure}

We considered \kepler{} targets observed in short cadence mode, which samples at an integration time $\Delta t=58.89$ s, or frequency $f_s=1/\Delta t=16980.8$ \muhz{}. There are about 5000 stars observed in this mode, and we detected about 50 subgiants. In this context, we define subgiants as stars showing oscillations of mixed modes, recognised as mode bumping (see Section~\ref{subsec:mode-iden} for a detailed discussion), which starts around the time of hydrogen exhaustion in the core. However, we required that the density of the mixed modes should be low enough, such that $\Delta\Pi_1>\Delta\nu/\nu^2$, where \dpi{1} is the period spacing of dipolar g modes. This occurs around the subgiant phase. Only subgiants with high signal-to-noise ratios (S/N) were selected to avoid any ambiguity of the correct assignment of the spherical degree $l$ of each mode (see Section~\ref{subsec:mode-iden}~below). Further, we required the observation duration, denoted by \tobs{}, to be at least two months. This resulted in a total of 36 subgiants that we analyse here.

The atmospheric parameters (\teff{} and [Fe/H]) were adopted from the KIC DR25 release \citep{mathur++2017-revised-properties-kepler-targets-dr25}. Table~\ref{tab:atm} lists the 36 stars in our sample along with \tobs{}, nicknames used in previous papers and references to any previous studies. Figure~\ref{fig:hrd} shows the targets in the \Dnu{}\,--\,\teff{} plane. They extend from the MS turnoff phase to the bottom of red giant branch. The MS stars in the \kepler{} LEGACY sample \citep{lund++2017-legacy-kepler-1} are also shown. 

We used light curves measured from simple aperture photometry (SAP) by the Kepler Science Center\footnote{https://archive.stsci.edu/kepler/}. We corrected instrumental effects with the \textsc{kasoc filter}\footnote{https://github.com/tasoc/corrections}. For a full description of this reduction procedure, see \citet{garcia++2011-kepler-lc-preparation} and \citet{handberg+2014-kepler-automated-time-seires-planet-host}. Briefly, it constructs moving-median high-pass filters, one with $\tau_{\rm long}=1/2$ d, and one with $\tau_{\rm short}=1/24$ d. Trends longer than $1/\tau_{\rm long}\sim23$ \muhz{} are removed, and any sharp feature that could produce drastically different signals for the two filters are eliminated. The light curves were converted into relative flux in ppm (parts per million) and we calculated power spectra by a Lomb-Scargle Periodogram algorithm \citep{1976Ap&SS..39..447L,1982ApJ...263..835S}, equivalent to a sine-wave fitting, as implemented in \textsc{astropy} \citep{2013A&A...558A..33A,2018arXiv180102634T}. The power density spectra were constructed by multiplying the power by the total observing time \citep{kjeldsen+1995-scaling-relations}. Figure~\ref{fig:gemma} uses Gemma (KIC 11026764) as an example to illustrate the power density spectrum and the spectral window. The former is comprised of a background and a group of peaks with a Gaussian-like envelope centred on \numax{}. Although there are small gaps in the original data, they have a negligible influence on the shape of oscillation modes. This can be seen from the fact that the spectral window has sidelobes with very low power.

\section{Parameter Estimation}
\label{sec:parameter}

\subsection{Mode identification}
\label{subsec:mode-iden}
\begin{figure}
	\includegraphics[width=\linewidth]{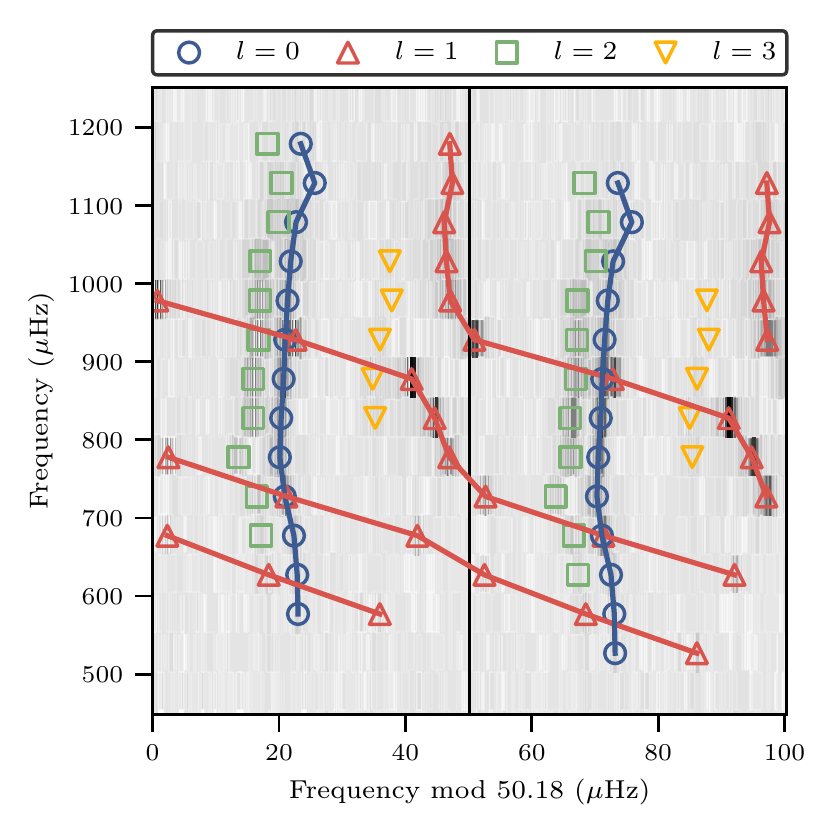}
    \caption{Replicated \'{e}chelle diagram for Gemma (KIC 11026764). By shifting one \Dnu{} downwards for the right replicate, the dipolar modes can be traced by a single line with one mode per horizontal row. The grey scale denotes the square-root of S/N to enhance contrast. The symbols represent extracted frequencies.}
    \label{fig:replicated-echelle}
\end{figure}

Each oscillation mode is associated with three quantum numbers $(n, l, m)$. The first step in mode identification is to assign an $l$-degree to each peak. In the asymptotic regime $(n\gg l)$ of p modes \citep{tassoul-1980-asymptotic-relation,gough-1986-aymptotic-relation}, this is straightforward because modes follow regular patterns. The frequencies are approximately equally spaced by \Dnu{}:
\beq \label{eq:pmode}
	\nu_{p}(n_p,l)\approx \Delta\nu\left( n_p+\frac{l}{2}+\epsilon_p \right) - \delta\nu_{0l}, 
\eeq
where \Dnu{} is the separation between adjacent radial modes, \epsp{} is the offset of the pattern, and the small separation \dnu{0l} defines the frequency differences between $l=0$ and $l$. The g modes, which we do not observe directly, are approximately equally spaced in period in the asymptotic regime:
\beq \label{eq:gmode}
	\Pi_{g}(n_g,l) = \nu_{g}^{-1}(n_g,l) \approx \Delta\Pi_{l}\left( n_g + \epsilon_g \right),
\eeq
where \dpi{l} and \epsg{} are the period spacing and offset for g modes. \citet{shibahashi-1979-modal-analysis-nonradial-oscillation-asymptotic-method} derived an asymptotic relation for mixed modes. Specifically, for dipolar modes ($l=1$) the mixed mode frequencies $\nu_m$ satisfy
\beq \label{eq:mmode}
	\tan \theta_p  = q\tan \theta_g ,
\eeq
where 
\beq
\label{eq:thetap}
\theta_p = \frac{\pi}{\Delta\nu}\left(\nu_m-\nu_p\right),
\eeq
\beq
\label{eq:thetag}
\theta_g = \frac{\pi}{\Delta\Pi_1}\left(\frac{1}{\nu_m}-\frac{1}{\nu_g}\right),
\eeq
and $q$ is a coupling factor determined by the width of the evanescent zone. Less-evolved subgiants have large period spacings, with one g mode often coupling with several p modes. The coupled modes can be bumped very far away from their regular p-mode spacings. By replicating \'{e}chelle diagrams horizontally, a line can trace all mixed modes, which helps mode identification \citep{2012ASPC..462..195B,benomar++2012-mass-sg-kepler}. See Figure~\ref{fig:replicated-echelle} for an illustration. Using this method, we were able to make unambiguous identifications for mixed dipolar modes, which still have a detectable amplitude even when they are bumped by a large amount. However, that is not the case for quadrupolar ($l=2$) modes due to their weaker couplings. Therefore, in most cases we only expect to detect one mixed quadrupolar mode in each order, even though sometimes two should be present. The same is true for octupolar ($l=3$) modes. To sum up, the dipolar mixed modes follow Equation~\ref{eq:mmode} closely, and the radial, quadrupolar and octupolar modes are distributed regularly, approximately as described by Equation~\ref{eq:pmode}.

We first obtained a crude estimate of \Dnu{} by tuning the horizontal width of the \echelle{} diagram such that the ridge of radial modes was vertical. To identify modes in each star, we smoothed the power spectrum and folded by \Dnu{} to create a collapsed power spectrum. This was cross-correlated with a template spectrum comprising three peaks, two narrower ($l=0$ and $l=2$) and one wider ($l=1$), to simulate the ridges. The value of $\epsilon_p$, which defines the location of $l=0$, could be correctly identified according to the lag at which the correlation coefficient was the largest. The modes satisfying $(\nu/\Delta\nu \mod 1) \in [\epsilon_p-0.04,\epsilon_p+0.04]$, $[\epsilon_p-0.12,\epsilon_p-0.04]$ and $[\epsilon_p+0.31,\epsilon_p+0.39]$ were labelled as $l=0$, $2$ and 3, respectively. These ranges were first empirically chosen. The reason is that \dnu{0l} approximately follows a linear trend with \Dnu{} and the slope is consistent with the ranges defined by those numbers (see Section 4). All other modes were provisionally assigned as $l=1$, noting that some $l=0,2$ and $3$ modes could actually be $l=1$. We used the relative height ($l=1$ modes are typically higher) and the regularity of those modes (Eq.~\ref{eq:pmode}) to help resolve the ambiguity. All labels were further confirmed in \'{e}chelle diagrams using the method mentioned above. We identified all modes within $\pm$ $8\Delta\nu$ around \numax{}. The identified peaks were used as inputs for the peakbagging.

\subsection{Power spectrum model}
We now describe the model used to fit the power spectrum. First, we consider the background caused by granulation, modelled as a sum of Lorentzian profiles:
\beq
	B(\nu) = \sum_{i=1}^{n}\frac{2\sqrt{2}}{\pi}\frac{a_i^2/b_i}{1+(\nu/b_i)^{c_i}}.
\eeq
The number of power profiles was set to be $n=1$ or $2$ \citep{harvey-1985-granulation,kallinger++2014-connection-granulation-oscillation}, subject to the goodness of fit, which is sufficient to account for the background near \numax{}. 
Second, we describe the signal of oscillations. 
For a mode with quantum number $(n,l,m)$, we adopted a power spectrum model
\beq
	L_{nlm}(\nu) = \frac{\mathcal{E}_{lm}(i^*)2A_{nl}^2/(\pi\Gamma_{nl})}{1+4(\nu-\nu_{nl}+m\nu_{s,nl})^2/\Gamma_{nl}^2}
\eeq
to account for oscillations, where the frequency $\nu_{nl}$, linewidth $\Gamma_{nl}$, amplitude $A_{nl}$, and rotational splitting $\nu_{s,nl}$ varies with $n$ and $l$ \citep{anderson++1990-solar-spectra-modelling}. The relative height within a multiplet depends on the inclination angle $i^*$ of the star and is modulated by a visibility function:
\beq
\mathcal{E}_{lm}(i^*) = \frac{(l-|m|)!}{(l+|m|)!}\left[P_l^{|m|}(\cos i^*)\right]^2,
\eeq
where $P_l^{|m|}$ are Legendre functions \citep{gizon+2003-determine-inclination-asteroseismology}.
The final model to fit the whole power spectrum was
\beq
\label{eq:Mmodel}
	M(\nu) = \eta^2(\nu)\left[\sum_{n=n_{\rm min}}^{n_{\rm max}} \sum_{l=0}^{3} \sum_{m=-l}^{l}L_{nlm}(\nu) + B(\nu) \right] + W.
\eeq
We used a frequency-independent constant $W$ to account for white/photon noise. The apodisation of the signal due to the integration time $\Delta t=58.89$ s was
\beq
	\eta^2(\nu) = \frac{\sin^2(\pi\nu\Delta t)}{(\pi\nu\Delta t)^2}.
\eeq
The free parameters to fit the power spectrum were $\theta = \{ \nu_{nl}, A_{nl}, \Gamma_{nl}, \nu_{s,nl}, i^*, a_i, b_i, c_i, W \}$.

\subsection{Fitting method}

We estimated the model parameters in the Bayesian framework \citep{handberg+2011-bayesian-peakbagging-mcmc-guide}. Given data $D$, model $M$ and prior information $I$, the parameters $\theta$ were estimated via posterior probability using Bayes' theorem:
\beq
\label{eq:bayes}
	p(\theta|D, M, I) = \frac{p(\theta|M, I) p(D|\theta, M, I)}{p(D|M, I)}.
\eeq
The posterior probability can be seen as a product of the prior and the likelihood function, divided by the Bayesian evidence, which is the marginalisation of that product over all parameter space.

We utilised three kinds of prior functions. The first was a uniform prior, which sets the prior probability as a constant over a parameter range:
\beq
	p(\theta|M,I) = \left\{ 
		\begin{aligned}
			& \frac{1}{\theta_{\text{max}}-\theta_{\text{min}}},\ \theta_{\text{min}}<\theta<\theta_{\text{max}}\\
			& 0, \ {\rm otherwise}\\
		\end{aligned}
		\right.
\eeq
We used uniform priors for most parameters. For example, the priors on the frequencies, $\nu_{nl}$, were set to be $3$\,\muhz{} ranges centred around their initial guessed values. The inclination angle was assigned a uniform prior across $[-\pi/2,\pi]$ and folded to $[0,\pi/2]$ after sampling. We also tested an isotropic prior for the inclination, but found no significant bias on mode parameters. The second is a modified Jeffreys prior, which was set for the linewidths and amplitudes:
\beq
	p(\theta|M,I) = \left\{
	\begin{aligned}
		& \frac{1}{\theta + \theta_{\rm uni}} \frac{1}{\log((\theta_{\rm uni}+\theta_{\rm max})/\theta_{\rm uni}) }, \ 0 < \theta < \theta_{\rm max} \\
		& 0, \ {\rm otherwise}\\
	\end{aligned}
	\right.
\eeq
We tested the impact of adopting uniform priors for these two particular parameters. The amplitudes do not present obvious bias while the linewidths are typically overestimated using the uniform priors. Despite the present bias, more than 95\% modes in our sample agree within 1-$\sigma$. 
For the splitting frequency, $\nu_{s,nl}$, we used the third prior, a flat prior with a half Gaussian: 
\beq
	p(\theta|M,I) = \left\{
	\begin{aligned}
		& 0, \ \theta < S\\
		& C, \ S\le \theta < U\\
		& C \cdot \exp{\left[-\frac{(\theta-U)^2}{2\sigma^2}\right]}, \ \theta\ge U\\
	\end{aligned}
	\right.
\eeq
where $C=(U-S+\sqrt{\pi/2}\sigma)^{-1}$. We applied this prior to $\nu_{s,nl}$ with $S=0$ \muhz{}, $U=1$ \muhz{} and $\sigma=0.5$ \muhz{} \citep{deheuvels++2014-internal-rotation-6-sg-rg-kepler}.


Gaussian noise in the time domain translates to a $\chi^2$ distribution with 2 degrees of freedom in the frequency domain (in power). Assuming all frequency bins, indexed by $i$ in the power spectrum, are statistically independent, the logarithm of the likelihood function is
\beq
	\ln p(D|\theta,M,I) = -\sum_{i}\left[ \ln M_i(\theta) + \frac{D_i}{M_i(\theta)} \right].
\eeq

\begin{figure}
	\includegraphics[width=\linewidth]{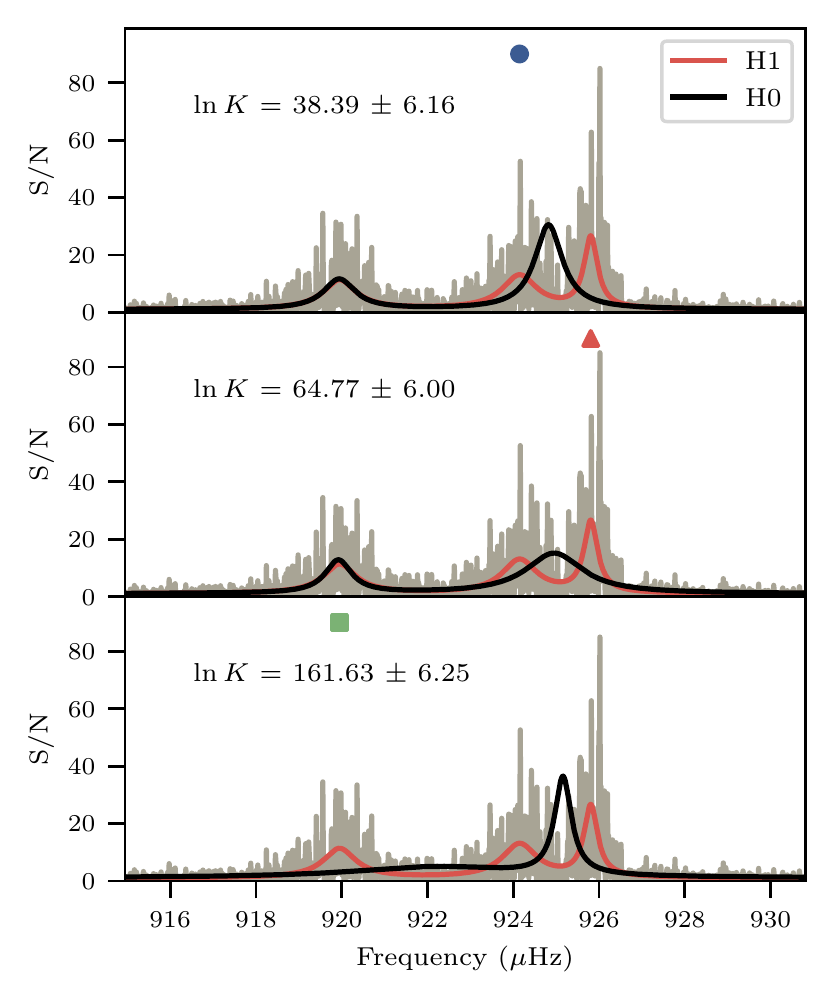}
    \caption{Significance tests for three modes of Gemma (KIC 11026764). Each panel presents the fits of the mode marked by the symbol, under H1 and H0 hypotheses separately. All three modes show very strong detections according to the \citet{kass&raftery} scale.}
    \label{fig:ortest}
\end{figure}

We used an H1 (odds ratio) approach to test whether a frequency range contains a mode \citep{appourchaux++2012-freq-ms-sg-kepler,corsaro+2014-diamonds,davies++2016-bayesian-peakbagging-35-kepler-planethost}. H1 is the hypothesis that a range of a power spectrum contains the mode, while H0 is the null hypothesis. 
In short, we computed the Bayes factor $\ln K=\ln p(D|M_1,I) - \ln p(D|M_0,I)$ and assessed it based on the \citet{kass&raftery} scale: 
\beq
\ln K = \left\{ \begin{array}{ll}
	<0  & \text{favours H0}  \\
	0-1 &  \text{not worth more than a bare mention}  \\
	1-3 & \text{positive}  \\
	3-5  & \text{strong} \\
	>5  & \text{very strong.}  
\end{array}\right.
\eeq

\subsection{Fitting details}
Based on the methods mentioned in the previous sections, we fitted the data as follows: 
\begin{enumerate}
    \item The power spectrum was initially fitted with Equation~\ref{eq:Mmodel} but the sum of Lorentzians was replaced by a single Gaussian profile centred around \numax{} (see Figure~\ref{fig:gemma}). The background spectrum was determined from this fit. By dividing the signal by the background, we obtained the S/N spectrum, which fluctuates around 1. 
    \item We selected seven modes with the highest amplitudes to fit simultaneously and to determine the inclination angle by maximising the posterior probability.
    \item The power spectrum was divided into segments and modes were fitted in each segment separately with a fixed inclination determined in step (ii). The sizes of the segments were based on the proximity of consecutive mode frequencies. Any two modes closer than 10 \muhz{} were grouped into the same segment. There was a minimum of one and a maximum of five modes per segment. The estimation of each parameter was obtained by marginalising the posterior probability and calculating median and 68 per cent credible limit values. 
    \item The Bayes factor was calculated for each mode to evaluate its significance, by comparing the Bayesian evidence with or without that mode in the model. Figure~\ref{fig:ortest} shows the fit of the three modes in a segment. All of them have very strong detections.
\end{enumerate}

We developed python software called \textsc{SolarlikePeakbagging}\footnote{https://github.com/parallelpro/SolarlikePeakbagging}, providing a wrapper for the Markov Chain Monte Carlo (MCMC) sampling algorithm implemented in \textsc{emcee} \citep{2010CAMCS...5...65G, 2013PASP..125..306F}, also featuring customisable Bayesian statistics, models, and other fitting algorithms. For the above fits, the sampler was chosen either to be an affine-invariant ensemble sampler (step ii) or a parallel-tempering sampler with 20 temperatures (step iii), depending on whether the Bayes factor was calculated. We initialised 500 walkers with values adopted from a least-squares fit, then burned-in for 1000 steps and iterated for 2000 steps. After each run, we checked convergence and the goodness of sampling by several metrics, including auto-correlation time, acceptance fraction, the evolution of model parameters, and the shapes of the marginal probability distributions.

\subsection{Fitting results}

\begin{figure*} 
\includegraphics[width=\linewidth]{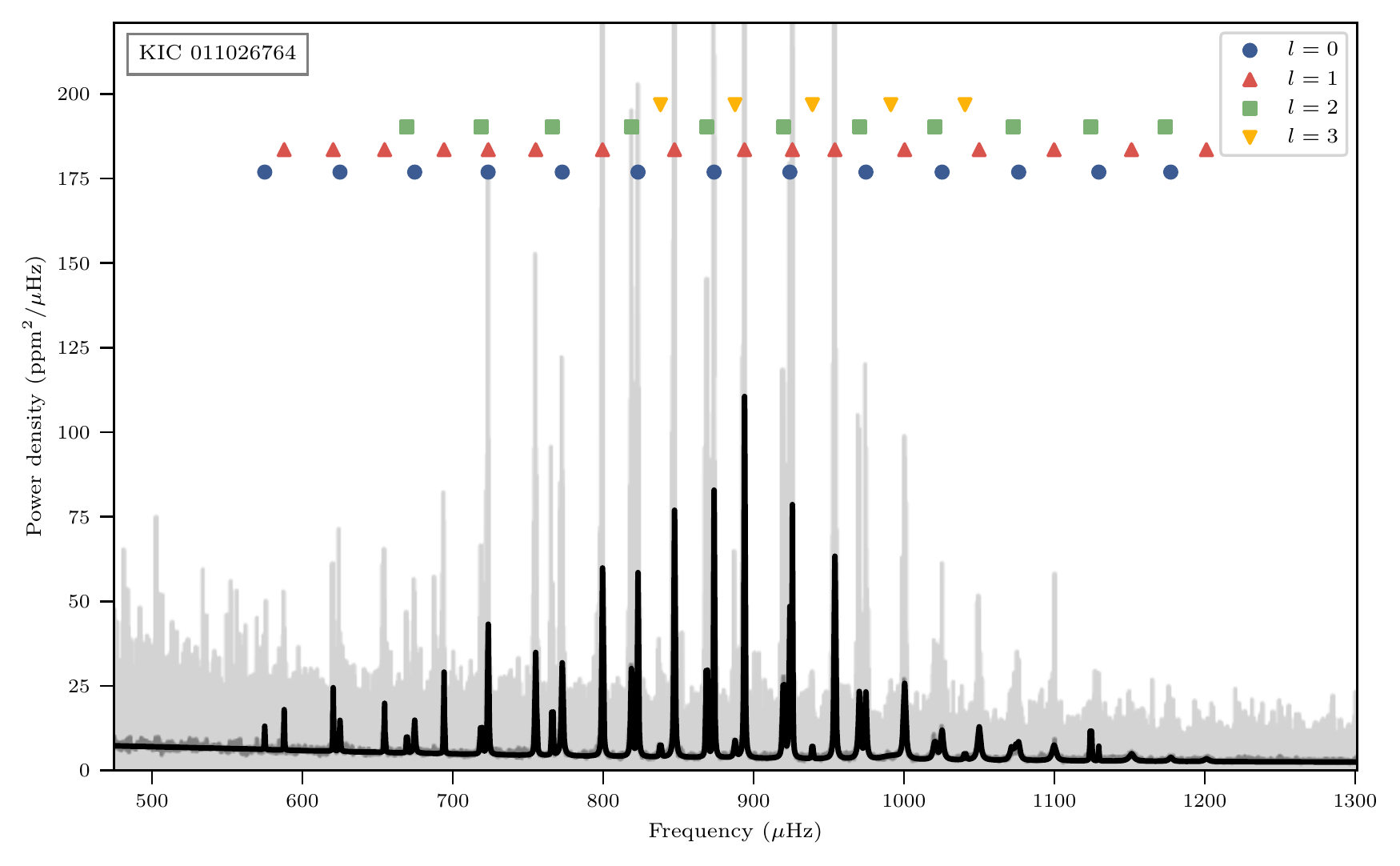} 
\caption{Power spectrum of Gemma (KIC 11026764). The fitted power spectrum (black) is overlaid on the original power spectrum (light grey).} 
\label{fig:gemma-ps} 
\end{figure*} 

\begin{table*} 
\caption{Mode parameters obtained from the peakbagging. These are $m=0$ modes. Only the first 10 rows are shown. The full table is available online.} 
\label{tab:mode-parameters-short} 
\begin{tabular*}{\textwidth}{@{\extracolsep{\fill}}rrrrrr}
\toprule
     KIC & $l$ &     $\nu_{nl}$ ($\mu$Hz) & $\Gamma_{nl}$ ($\mu$Hz) &        $A_{nl}$ (ppm) &               $\ln K$ \\
\midrule
 2991448 &   0 &   $889.34$ $\pm$ $0.312$ &     $0.79$ $\pm$ $0.96$ &   $5.72$ $\pm$ $1.53$ &   $2.16$ $\pm$ $0.75$ \\
 2991448 &   0 &   $949.46$ $\pm$ $1.111$ &     $8.02$ $\pm$ $2.37$ &  $12.95$ $\pm$ $1.61$ &   $3.61$ $\pm$ $0.98$ \\
 2991448 &   0 &  $1011.17$ $\pm$ $0.202$ &     $1.15$ $\pm$ $0.46$ &   $9.47$ $\pm$ $1.23$ &  $20.49$ $\pm$ $4.32$ \\
 2991448 &   0 &  $1072.78$ $\pm$ $0.099$ &     $0.76$ $\pm$ $0.24$ &  $14.78$ $\pm$ $1.49$ &  $39.42$ $\pm$ $2.54$ \\
 2991448 &   0 &  $1134.18$ $\pm$ $0.113$ &     $0.89$ $\pm$ $0.28$ &  $14.21$ $\pm$ $1.38$ &  $47.86$ $\pm$ $2.81$ \\
 2991448 &   0 &  $1195.89$ $\pm$ $0.547$ &     $3.29$ $\pm$ $1.72$ &  $13.37$ $\pm$ $2.38$ &   $5.40$ $\pm$ $2.13$ \\
 2991448 &   0 &  $1257.81$ $\pm$ $0.969$ &     $3.53$ $\pm$ $2.63$ &  $10.04$ $\pm$ $2.96$ &   $2.68$ $\pm$ $1.52$ \\
 2991448 &   0 &  $1319.52$ $\pm$ $1.294$ &     $3.00$ $\pm$ $3.56$ &   $5.09$ $\pm$ $3.29$ &   $0.18$ $\pm$ $1.20$ \\
 2991448 &   1 &   $857.88$ $\pm$ $0.384$ &     $0.97$ $\pm$ $0.79$ &   $5.88$ $\pm$ $1.62$ &  $-0.21$ $\pm$ $0.69$ \\
 2991448 &   1 &   $915.60$ $\pm$ $0.153$ &     $0.81$ $\pm$ $0.32$ &   $9.62$ $\pm$ $1.27$ &  $19.19$ $\pm$ $1.16$ \\
\bottomrule
\end{tabular*}
\end{table*}

\begin{figure*}
\includegraphics[width=\linewidth]{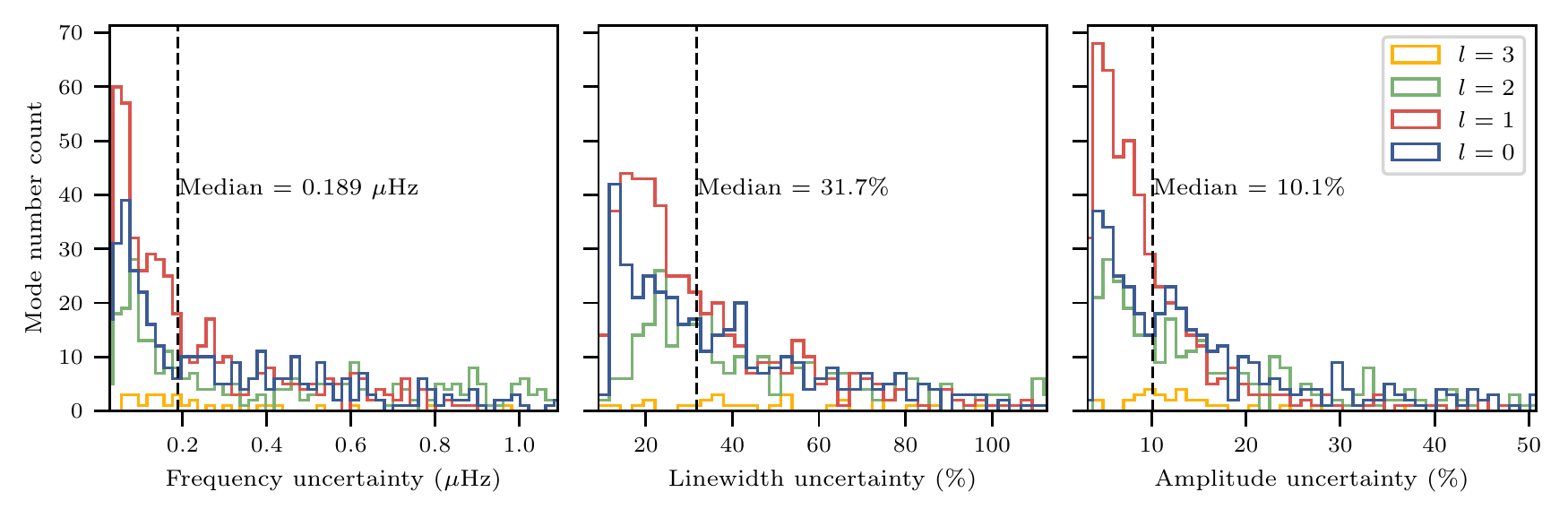}
\caption{Histograms of the uncertainties for measured mode frequencies (left), linewidths (middle) and amplitudes (right). The linewidths and amplitudes are shown in relative uncertainty.}
\label{fig:uncertainty-hist} 
\end{figure*}


We present mode frequencies, amplitudes and linewidths for $m=0$ modes in Table~\ref{tab:mode-parameters-short}. The rotational splittings will be comprehensively studied in a future paper. Figure~\ref{fig:gemma-ps} shows the power spectrum of Gemma (KIC 11026764) with fitted mode frequencies overlaid. We present the plots of the other stars in the Appendix. 

In Figure~\ref{fig:uncertainty-hist}, we show the histograms of uncertainties. The median value of frequency uncertainties is 0.180 \muhz{}. As expected, the uncertainty is a function of S/N. The typical uncertainty is 0.1 \muhz{} for S/N=3. The median uncertainties of linewidths are 31.7\% and amplitudes are 10.1\%. 

As a check, we can consider the classical maximum likelihood estimator \citep[MLE;][]{libbrecht-1992-accuracy-solar-frequency-measures,toutain+1994-mle,ballot++2008-spectra-fitting-inclined-stars}. This calculates uncertainties by inverting the Hessian matrix whose elements are the second derivatives of the likelihood function to the parameters. The Cram\'{e}r-Rao bound states that an unbiased MLE reaches the lowest variance bound, so any other unbiased estimators are expected to obtain a larger variance. \citet{libbrecht-1992-accuracy-solar-frequency-measures} derived an analytical form of the uncertainty. 
We found the estimated uncertainties derived from the MCMC-based posterior distributions were typically larger than those obtained from the above analytical forms, by factors of 1.15 (frequency), 1.18 (linewidth) and 1.41 (height), respectively. Thus the uncertainties are safe to use for modelling.

\begin{figure}
	\includegraphics[width=\linewidth]{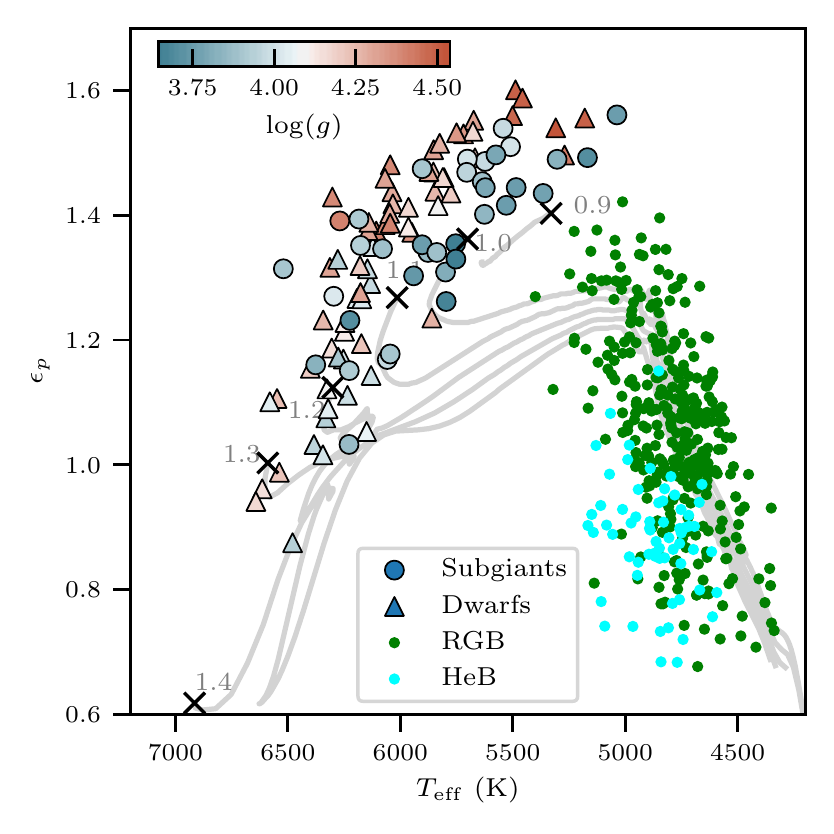}
    \caption{\epsp{} vs. \teff{} with colour-coded \logg{}. The measurement of red giants were adopted from \citet{huber++2010-asteroseismology-800-rg-kepler-global-parameters}. The theoretical evolutionary tracks \citep{white++2011-asteroseismic-diagrams-cd-epsilon-deltaP-models} are labelled with mass in solar units. The crosses mark the zero-age MS.}
    \label{fig:epsp}
\end{figure}


\begin{figure}
	\includegraphics[width=\linewidth]{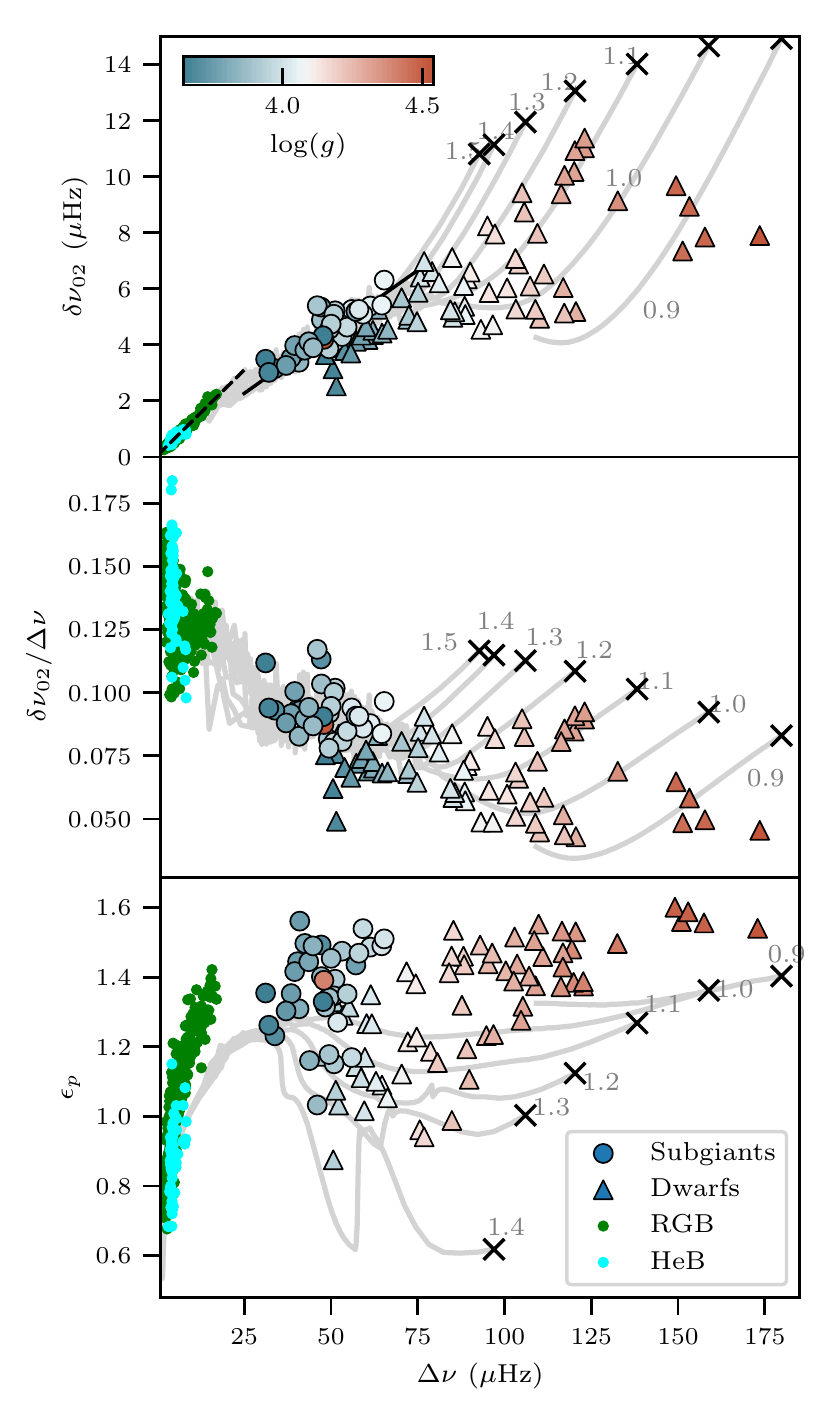}
    \caption{Top and middle: C\,--\,D diagrams, which show the small separation $\delta\nu_{02}$ as a function of \Dnu{}, with colour-coded \logg{}. The solid line is a linear fit to the subgiants, and the dashed line is a similar fit to low-luminosity red giants \citep{bedding++2010-kepler-rg}. Bottom: \epsp{} vs. \Dnu{}. The measurement of dwarfs and red giants were adopted from \citet{lund++2017-legacy-kepler-1} and \citet{huber++2010-asteroseismology-800-rg-kepler-global-parameters}, respectively. The theoretical evolutionary tracks \citep{white++2011-asteroseismic-diagrams-cd-epsilon-deltaP-models} are labelled with mass in solar units with crosses marking the zero-age MS.}
    \label{fig:c-d-diagram}
\end{figure}

\begin{figure}
	\includegraphics[width=\linewidth]{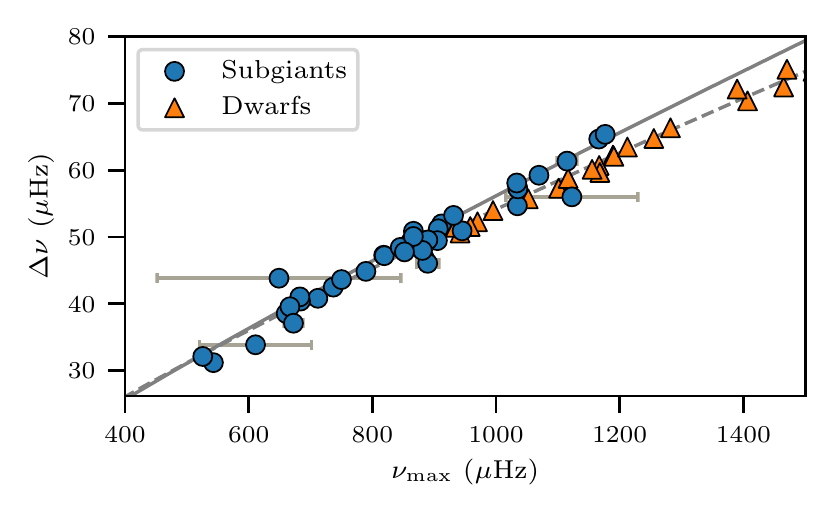}
    \caption{\Dnu{}\,--\,\numax{} diagram. The solid line is the best-fitted curve for subgiants. The dashed line is a similar fit, but obtained from a sample of MS dwarfs and red giants \citep{huber++2011-test-scaling-relation-ms-rgb-kepler}. The measurement of dwarfs were adopted from \citet{lund++2017-legacy-kepler-1}.}
    \label{fig:dnu_vs_numax}
\end{figure}

\section{Mode Frequencies}
\label{sec:freq}

\begin{table*} 
\caption{Global Oscillation parameters.} 
\label{tab:dnu} 
\begin{tabular*}{\textwidth}{@{\extracolsep{\fill}}rrrr p{1.0cm} p{1.0cm}r rrrr}
\toprule
      KIC & $\Delta\nu$ ($\mu$Hz) &           $\epsilon_p$ & $\nu_\text{max}$ ($\mu$Hz) &  $A_\text{max}$ (ppm) &        Width ($\mu$Hz) & $\delta\nu_{02}$ ($\mu$Hz) & $\delta\nu_{03}$ ($\mu$Hz) &   $\gamma_1$ ($\mu$Hz) & $\Delta\Pi_1$ (s) \\
\midrule
  2991448 &  $61.35$ $\pm$ $0.05$ &  $1.486$ $\pm$ $0.013$ &          $1115$ $\pm$ $16$ &   $9.83$ $\pm$ $0.45$ &   $187.3$ $\pm$ $21.9$ &        $5.38$ $\pm$ $0.41$ &                        --- &   $1022.0$ $\pm$ $0.3$ &               --- \\
  3852594 &  $51.95$ $\pm$ $0.06$ &  $1.270$ $\pm$ $0.018$ &            $912$ $\pm$ $8$ &  $13.00$ $\pm$ $0.35$ &   $181.3$ $\pm$ $12.7$ &                        --- &                        --- &                    --- &               --- \\
  4346201 &  $56.00$ $\pm$ $0.04$ &  $1.169$ $\pm$ $0.014$ &         $1123$ $\pm$ $107$ &   $8.07$ $\pm$ $0.48$ &  $366.5$ $\pm$ $129.1$ &        $5.26$ $\pm$ $0.53$ &                        --- &    $812.2$ $\pm$ $2.6$ &               --- \\
  5108214 &  $40.81$ $\pm$ $0.02$ &  $1.309$ $\pm$ $0.007$ &            $712$ $\pm$ $7$ &   $9.51$ $\pm$ $0.35$ &   $197.2$ $\pm$ $11.8$ &        $3.38$ $\pm$ $0.24$ &                        --- &                    --- &  $411$ $\pm$ $23$ \\
  5607242 &  $40.39$ $\pm$ $0.01$ &  $1.444$ $\pm$ $0.003$ &            $684$ $\pm$ $2$ &  $10.76$ $\pm$ $0.27$ &    $114.4$ $\pm$ $3.7$ &        $3.77$ $\pm$ $0.14$ &        $8.77$ $\pm$ $0.13$ &                    --- &  $176$ $\pm$ $16$ \\
  5689820 &  $41.02$ $\pm$ $0.01$ &  $1.561$ $\pm$ $0.002$ &            $683$ $\pm$ $4$ &  $13.85$ $\pm$ $0.47$ &    $100.1$ $\pm$ $5.0$ &        $3.81$ $\pm$ $0.21$ &        $9.24$ $\pm$ $0.39$ &                    --- &   $154$ $\pm$ $2$ \\
  5955122 &  $49.22$ $\pm$ $0.01$ &  $1.340$ $\pm$ $0.004$ &            $861$ $\pm$ $4$ &   $9.04$ $\pm$ $0.14$ &    $175.5$ $\pm$ $5.6$ &        $4.02$ $\pm$ $0.23$ &                        --- &  $1143.9$ $\pm$ $10.0$ &               --- \\
  6064910 &  $43.82$ $\pm$ $0.08$ &  $1.160$ $\pm$ $0.032$ &          $649$ $\pm$ $197$ &  $12.75$ $\pm$ $0.98$ &  $456.8$ $\pm$ $275.9$ &                        --- &                        --- &                    --- &   $390$ $\pm$ $5$ \\
  6370489 &  $51.22$ $\pm$ $0.03$ &  $1.394$ $\pm$ $0.010$ &           $907$ $\pm$ $13$ &   $9.50$ $\pm$ $0.44$ &   $220.6$ $\pm$ $19.7$ &        $5.21$ $\pm$ $0.37$ &                        --- &    $937.6$ $\pm$ $0.8$ &               --- \\
  6442183 &  $64.65$ $\pm$ $0.01$ &  $1.490$ $\pm$ $0.002$ &           $1166$ $\pm$ $3$ &   $7.82$ $\pm$ $0.13$ &    $206.1$ $\pm$ $4.3$ &        $5.42$ $\pm$ $0.21$ &       $12.33$ $\pm$ $0.09$ &   $1005.6$ $\pm$ $0.1$ &               --- \\
  6693861 &  $47.26$ $\pm$ $0.01$ &  $1.402$ $\pm$ $0.005$ &            $818$ $\pm$ $9$ &  $10.76$ $\pm$ $0.38$ &   $147.9$ $\pm$ $12.3$ &        $4.89$ $\pm$ $0.13$ &                        --- &                    --- &               --- \\
  6766513 &  $50.91$ $\pm$ $0.04$ &  $1.151$ $\pm$ $0.013$ &           $945$ $\pm$ $18$ &   $9.03$ $\pm$ $0.43$ &   $288.3$ $\pm$ $29.2$ &        $4.09$ $\pm$ $0.50$ &                        --- &    $799.3$ $\pm$ $0.4$ &               --- \\
  7174707 &  $47.19$ $\pm$ $0.01$ &  $1.492$ $\pm$ $0.003$ &            $819$ $\pm$ $4$ &  $11.22$ $\pm$ $0.30$ &    $118.4$ $\pm$ $4.9$ &        $5.35$ $\pm$ $0.23$ &       $10.14$ $\pm$ $0.18$ &                    --- &   $172$ $\pm$ $5$ \\
  7199397 &  $38.54$ $\pm$ $0.01$ &  $1.353$ $\pm$ $0.005$ &            $661$ $\pm$ $6$ &   $9.87$ $\pm$ $0.21$ &    $148.9$ $\pm$ $6.5$ &        $3.53$ $\pm$ $0.25$ &                        --- &                    --- &  $228$ $\pm$ $45$ \\
  7668623 &  $46.05$ $\pm$ $0.05$ &  $1.033$ $\pm$ $0.019$ &           $889$ $\pm$ $18$ &   $8.77$ $\pm$ $0.39$ &   $216.8$ $\pm$ $25.4$ &        $5.40$ $\pm$ $0.60$ &                        --- &    $675.2$ $\pm$ $1.1$ &               --- \\
  7747078 &  $53.22$ $\pm$ $0.01$ &  $1.475$ $\pm$ $0.002$ &            $931$ $\pm$ $5$ &   $8.14$ $\pm$ $0.13$ &    $187.7$ $\pm$ $5.8$ &        $4.30$ $\pm$ $0.24$ &                        --- &   $1031.7$ $\pm$ $0.2$ &               --- \\
  7976303 &  $50.85$ $\pm$ $0.01$ &  $1.346$ $\pm$ $0.003$ &            $866$ $\pm$ $4$ &   $9.21$ $\pm$ $0.13$ &    $182.3$ $\pm$ $4.7$ &        $5.09$ $\pm$ $0.22$ &                        --- &    $979.4$ $\pm$ $0.6$ &               --- \\
  8026226 &  $33.85$ $\pm$ $0.02$ &  $1.231$ $\pm$ $0.010$ &           $611$ $\pm$ $90$ &   $8.03$ $\pm$ $0.54$ &  $334.5$ $\pm$ $160.0$ &        $3.15$ $\pm$ $0.31$ &                        --- &                    --- &  $343$ $\pm$ $25$ \\
  8524425 &  $59.23$ $\pm$ $0.01$ &  $1.539$ $\pm$ $0.001$ &           $1069$ $\pm$ $3$ &   $8.96$ $\pm$ $0.15$ &    $184.2$ $\pm$ $4.4$ &        $5.09$ $\pm$ $0.22$ &       $10.15$ $\pm$ $0.72$ &   $1049.6$ $\pm$ $0.1$ &               --- \\
  8702606 &  $39.55$ $\pm$ $0.01$ &  $1.416$ $\pm$ $0.002$ &            $667$ $\pm$ $2$ &  $10.74$ $\pm$ $0.20$ &    $116.8$ $\pm$ $2.5$ &        $3.97$ $\pm$ $0.08$ &        $8.70$ $\pm$ $0.25$ &                    --- &  $167$ $\pm$ $17$ \\
  8738809 &  $49.44$ $\pm$ $0.02$ &  $1.178$ $\pm$ $0.008$ &            $905$ $\pm$ $9$ &   $8.40$ $\pm$ $0.31$ &   $217.6$ $\pm$ $11.4$ &        $3.86$ $\pm$ $0.24$ &                        --- &    $852.4$ $\pm$ $0.2$ &               --- \\
  9512063 &  $49.55$ $\pm$ $0.02$ &  $1.340$ $\pm$ $0.008$ &           $890$ $\pm$ $23$ &   $7.93$ $\pm$ $0.53$ &   $238.1$ $\pm$ $62.7$ &        $4.55$ $\pm$ $0.27$ &                        --- &                    --- &               --- \\
 10018963 &  $54.66$ $\pm$ $0.01$ &  $1.352$ $\pm$ $0.003$ &           $1035$ $\pm$ $6$ &   $7.55$ $\pm$ $0.12$ &    $246.5$ $\pm$ $5.9$ &        $4.63$ $\pm$ $0.19$ &                        --- &    $786.0$ $\pm$ $0.3$ &               --- \\
 10147635 &  $37.08$ $\pm$ $0.01$ &  $1.303$ $\pm$ $0.006$ &           $672$ $\pm$ $15$ &   $8.84$ $\pm$ $0.39$ &   $190.9$ $\pm$ $19.6$ &        $3.26$ $\pm$ $0.25$ &                        --- &    $683.9$ $\pm$ $0.3$ &               --- \\
 10273246 &  $47.98$ $\pm$ $0.02$ &  $1.391$ $\pm$ $0.006$ &            $881$ $\pm$ $8$ &  $10.58$ $\pm$ $0.36$ &   $189.8$ $\pm$ $10.4$ &        $4.19$ $\pm$ $0.28$ &                        --- &    $977.8$ $\pm$ $1.0$ &               --- \\
 10593351 &  $31.18$ $\pm$ $0.01$ &  $1.354$ $\pm$ $0.006$ &            $543$ $\pm$ $5$ &  $12.77$ $\pm$ $0.47$ &    $133.8$ $\pm$ $7.1$ &        $3.49$ $\pm$ $0.24$ &                        --- &                    --- &  $350$ $\pm$ $23$ \\
 10873176 &  $48.43$ $\pm$ $0.11$ &  $1.314$ $\pm$ $0.037$ &           $845$ $\pm$ $11$ &  $14.34$ $\pm$ $0.90$ &   $117.6$ $\pm$ $11.8$ &                        --- &                        --- &                    --- &               --- \\
 10920273 &  $57.19$ $\pm$ $0.02$ &  $1.435$ $\pm$ $0.007$ &          $1035$ $\pm$ $13$ &  $10.73$ $\pm$ $0.40$ &   $185.4$ $\pm$ $22.0$ &        $5.19$ $\pm$ $0.22$ &                        --- &                    --- &               --- \\
 10972873 &  $58.07$ $\pm$ $0.01$ &  $1.469$ $\pm$ $0.003$ &           $1033$ $\pm$ $5$ &   $8.84$ $\pm$ $0.17$ &    $186.3$ $\pm$ $6.1$ &        $5.26$ $\pm$ $0.21$ &       $10.57$ $\pm$ $0.13$ &   $1041.7$ $\pm$ $0.1$ &               --- \\
 11026764 &  $50.07$ $\pm$ $0.01$ &  $1.454$ $\pm$ $0.003$ &            $866$ $\pm$ $3$ &   $8.08$ $\pm$ $0.13$ &    $171.5$ $\pm$ $4.0$ &        $4.74$ $\pm$ $0.19$ &       $10.09$ $\pm$ $0.23$ &    $926.4$ $\pm$ $0.1$ &               --- \\
 11137075 &  $65.35$ $\pm$ $0.01$ &  $1.510$ $\pm$ $0.003$ &           $1177$ $\pm$ $7$ &   $9.66$ $\pm$ $0.29$ &   $181.7$ $\pm$ $11.0$ &        $6.31$ $\pm$ $0.18$ &                        --- &   $1143.6$ $\pm$ $0.1$ &               --- \\
 11193681 &  $42.48$ $\pm$ $0.01$ &  $1.497$ $\pm$ $0.004$ &            $737$ $\pm$ $4$ &  $10.21$ $\pm$ $0.20$ &    $151.5$ $\pm$ $5.6$ &        $3.81$ $\pm$ $0.16$ &                        --- &                    --- &               --- \\
 11395018 &  $47.76$ $\pm$ $0.02$ &  $1.330$ $\pm$ $0.006$ &           $852$ $\pm$ $10$ &  $10.35$ $\pm$ $0.25$ &   $171.7$ $\pm$ $15.9$ &        $4.32$ $\pm$ $0.18$ &                        --- &                    --- &               --- \\
 11414712 &  $43.62$ $\pm$ $0.01$ &  $1.444$ $\pm$ $0.002$ &            $750$ $\pm$ $3$ &   $8.73$ $\pm$ $0.15$ &    $146.4$ $\pm$ $3.4$ &        $4.11$ $\pm$ $0.13$ &       $10.07$ $\pm$ $0.20$ &                    --- &  $204$ $\pm$ $17$ \\
 11771760 &  $32.10$ $\pm$ $0.02$ &  $1.262$ $\pm$ $0.008$ &            $526$ $\pm$ $5$ &  $12.20$ $\pm$ $0.35$ &    $106.7$ $\pm$ $7.7$ &        $3.02$ $\pm$ $0.13$ &                        --- &                    --- &  $191$ $\pm$ $10$ \\
 12508433 &  $44.84$ $\pm$ $0.00$ &  $1.490$ $\pm$ $0.002$ &            $790$ $\pm$ $2$ &  $11.14$ $\pm$ $0.21$ &    $111.7$ $\pm$ $2.0$ &        $3.89$ $\pm$ $0.17$ &        $9.32$ $\pm$ $0.23$ &                    --- &   $164$ $\pm$ $9$ \\
\bottomrule
\end{tabular*}
\end{table*} 


We fitted the radial mode frequencies with Equation~\ref{eq:pmode} to estimate \Dnu{} and \epsp{}. The errors were considered within the Bayesian framework thus they were propagated from the priors and likelihoods. The likelihood function was selected to reflect the residuals between data and the models, weighted by uncorrelated uncertainties from both dependent and independent variables, as normal distributions. 

The p mode offset \epsp{} is sensitive to the lower and upper turning points of a mode \citep[e.g.][]{gough-1986-aymptotic-relation}. It has been demonstrated to vary with \teff{} and \Dnu{} \citep{white++2011-asteroseismic-diagrams-cd-epsilon-deltaP-models,hon++2018-deep-learning-15000kepler-rg-k2-tess}, and was used to resolve ambiguous mode identification problems found in F-type dwarfs \citep{white++2012-mode-identification-bloody-f-stars} and to discriminate between red-giant-branch (RGB) and helium-core-burning (HeB) stars \citep{kallinger++2012-epsp}. The upper turning point lies at the surface, where \teff{} and \logg{} are relevant quantities, so it may be conjectured that \epsp{} is related to both of them. In Figure~\ref{fig:epsp} we present \epsp{} as a function of \teff{}. As \citet{ong+2019-model-epsp} pointed out, \epsp{} depends on the method of measurement, so we recalculated the \epsp{} of the MS dwarfs from \citet{lund++2017-legacy-kepler-1} using our approach. As the figure shows, the general trend of \epsp{} in subgiants follows a similar pattern as for dwarfs. We also identify a dependence on \logg{}: stars with similar \teff{} but smaller \logg{} have lower \epsp{}. The dwarf stars, which have higher \logg{}, sit above the subgiants on the \epsp{}\,--\,\teff{} diagram. In the bottom panel of Figure~\ref{fig:c-d-diagram}, we present \epsp{} as a function of \Dnu{}. The obvious offset between observations and stellar models stems from the improper modelling of near-surface layers \citep{jcd++1988-solar-freq}.

We show the small separations \dnu{02} versus \Dnu{} in Figure~\ref{fig:c-d-diagram}, the so-called C\,--\,D diagram \citep{jcd-1984-review,mazumdar-2005-c-d-diagram}. The evolutionary tracks with different masses are well separated on the main sequence. By comparing to models, the diagram is useful for estimating mass and age when \dnu{02} and \Dnu{} are known. We note that the \epsp{}\,--\,\Dnu{} diagram could be used for a similar purpose, provided the surface effect is well accounted for. However, the tracks become more degenerate in the subgiant phase \citep{white++2011-asteroseismic-diagrams-76-solartype-stars,white++2011-asteroseismic-diagrams-cd-epsilon-deltaP-models}. We performed a linear fit to the subgiants:
\beq
	\delta\nu_{02} = (0.088\pm0.001)\Delta\nu + (0.070\pm0.059) \ \mu\text{Hz}.
\eeq
We point out that some obvious outliers from the fitted line are present. They are affected by bumped $l=2$ modes, which lift the average distance of a quadrupolar mode to a radial mode. \citet{bedding++2010-kepler-rg} fitted the same relation to a sample of red giants, which have a steeper slope (as shown by the dashed line in Figure~\ref{fig:c-d-diagram}). This difference is also visible from the calculation of stellar models. Similarly, we fitted $\delta\nu_{03}$ using ten stars that have detected $l=3$ modes:
\beq
	\delta\nu_{03} = (0.152\pm0.002)\Delta\nu + (2.578\pm0.102) \ \mu\text{Hz}.
\eeq


Finally, we use the radial modes to determine \numax{} because they are purely acoustic, since there are no radial g modes for them to couple with. By fitting the amplitudes versus frequencies of radial modes with a Gaussian function, \numax{} was obtained as the Gaussian's centre. The height of the Gaussian, \amax{}, together with the width, are analysed in Section~\ref{sec:amp}. We fitted the well-established $\Delta\nu-\nu_\text{max}$ relation \citep[e.g.][]{stello++2009-relation-dnu-numax,huber++2009-syd-pipeline} to our subgiants:
\beq\label{eq:dnu-numax}
	\Delta\nu = a(\nu_\text{max}/{\mu\text{Hz}})^b \ \mu\text{Hz},
\eeq
and determined the best fitted parameters as $a=0.158\pm0.036$ and $b=0.847\pm0.034$. \citet{huber++2011-test-scaling-relation-ms-rgb-kepler} determined $a=0.22$ and $b=0.797$ from a sample consisting of both MS stars and red giants. We plot the two relations in Figure~\ref{fig:dnu_vs_numax}. The most noticeable effect is that the subgiants lie above the dwarfs, which can be understood from simple scaling arguments. The asteroseismic scaling relations \citep{brown++1991-dection-procyon-scaling-relation,kjeldsen+1995-scaling-relations} state that $\Delta\nu\propto\sqrt{\rho}\propto M^{1/2}R^{-3/2}$, and $\nu_{\rm max}\propto g/\sqrt{T_{\rm eff}}\propto MR^{-2}T_{\rm eff}^{-1/2}$. Dividing the two relations, we obtain $\Delta\nu/\nu_{\rm max} \propto M^{-1/2}R^{1/2}T_{\rm eff}^{-1/2}$. The subgiants and dwarfs have similar $M$ and \teff{}, but subgiants are more inflated, which ultimately leads to higher $\Delta\nu/\nu_{\rm max}$. Indeed, this spread is the signal upon which the widely-used scaling relations for mass and radius are based \citep{stello++2008-wire-kgiants,kallinger++2010-rg-corot-mass-radius}.


\section{The p--g diagram}
\label{sec:p-g-diagram}

\begin{figure}
	\includegraphics[width=\linewidth]{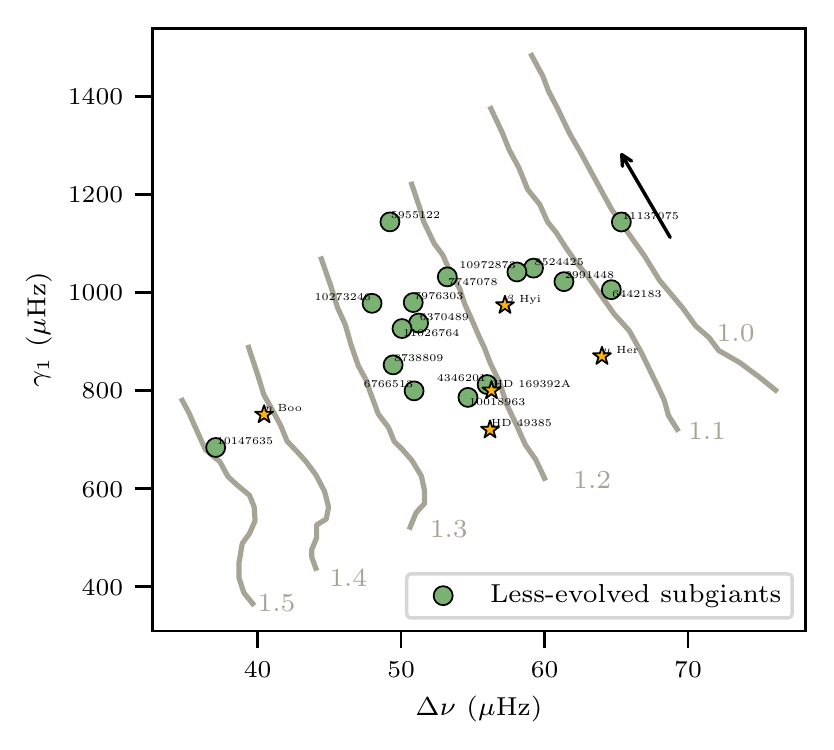}
    \caption{The p--g diagram using the lowest-order dipolar g mode \nug{1} vs. \Dnu{} for less-evolved subgiants. We also include five stars observed by ground-based telescopes and \corot{}, shown in the star symbols. The theoretical evolutionary tracks (Paper II) are labelled with mass in solar units, with the black arrow pointing in the direction of evolution.}
    \label{fig:gamma1_vs_dnu}
\end{figure}

\begin{figure}
	\includegraphics[width=\linewidth]{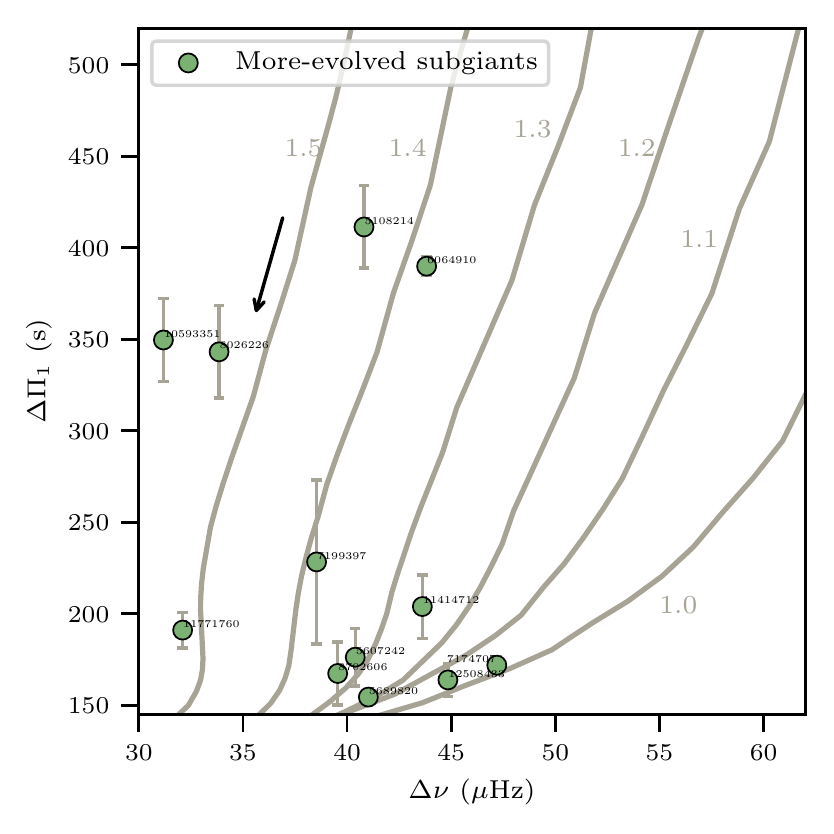}
    \caption{The dipolar g mode period spacing \dpi{1} vs. \Dnu{} for more-evolved subgiants. The theoretical evolutionary tracks \citep{Choi++2016-mist-1-solar-scaled-models} are labelled with mass in solar units, with the black arrow pointing in the direction of evolution.}
    \label{fig:dpi1_vs_dnu}
\end{figure}

We now focus on the frequencies of the underlying g modes, which we denote by \nug{} \citep{aizenman++1977-avoided-crossing}. Adjusting the model parameters to fit the observed frequencies can be difficult and time-consuming. As pointed out by \citet{bedding-2014-solar-like-review}, the information contained in the mixed modes can be used more elegantly, by considering the frequencies of the avoided crossings themselves. The suggestion was that, before fitting models to the observed mixed modes, one can first consider the underlying g modes. These are not directly detected but their frequencies can be inferred from the locations of the avoided crossings. 

This discussion suggests a new asteroseismic diagram \citep{bedding-2014-solar-like-review}, inspired by the classical C--D diagram, in which the frequencies of the avoided crossings \nug{} are plotted against the large separation of the p modes. This p--g diagram, so named because it plots g mode frequencies versus p mode frequencies, could prove to be an instructive way to display results of many stars and to make a first comparison with theoretical models \citep{bedding-2014-solar-like-review,campante++2011-subgiants-kic10273246-kic10920273}.

For example, the \'{e}chelle diagram in Figure~\ref{fig:replicated-echelle} shows three avoided crossings of dipole modes, at about 950, 750 and 650 \muhz{}. We recognise these as the frequencies of the \nug{} modes, that is, the pure g~modes that would exist in the core cavity if there was no coupling to the p modes in the envelope \citep{aizenman++1977-avoided-crossing}. Much of the diagnostic information contained in the mixed modes can be captured in this way. This is because the overall pattern of the mixed modes is determined by the mode bumping at each avoided crossing, and these patterns are determined by the g modes trapped in the core.  For related discussion on this point, see \citet{deheuvels+2010-hd49385-corot-interior-model} and \citet{benomar++2013-avoided-crossing-sg-kepler}.

We determined the g mode frequencies, $\gamma$, from fitting $l=1$ mixed mode frequencies $\nu_{n, l=1}$. Specifically, Lorentzian profiles were fitted to $\nu_{n+1, l=1}-\nu_{n, l=1}$ versus $(\nu_{n+1, l=1}+\nu_{n, l=1})/2$. A dip is present wherever there is an avoided crossing. This is because the presence of mixed modes makes two adjacent modes closer in frequency. We identified the centres of the Lorentzian profiles as the g mode frequencies, $\gamma$. The errors were estimated following a Monte Carlo procedure by adding Gaussian errors to mode frequencies, repeating the fitting for 1000 times, and adopting the standard deviation of $\gamma$ from the distribution. If multiple avoided crossings (at least three) are present, we further estimated \dpi{1} by calculating the average differences between $1/\gamma$. The results are shown in Table~\ref{tab:dnu}. Some stars do not have reported \nug{1} or \dpi{1} because they tend to have uncertain identifications of the first g mode, and they also present a small number of avoided crossings.

For less-evolved subgiants in which the first ($n_g=1$) g mode frequency, \nug{1}, is present, we made the \nug{1}\,--\,\Dnu{} diagram in Figure~\ref{fig:gamma1_vs_dnu}. This provides an indicative measurement of stellar mass when evolutionary tracks are simultaneously plotted, unlike the C--D diagram where stellar tracks are degenerate. The stars evolve diagonally, so \nug{1} and \Dnu{} can determine not only mass, but age as well. 

Another important diagnostic diagram is to plot the period spacings of g modes \dpi{1} in Equation~\ref{eq:gmode} versus \Dnu{} (another p\,--\,g diagram). The diagram has been proved useful for distinguishing RGB and HeB stars because the two types of stars cluster at different locations \citep{bedding++2011-distinguish-rc-rgb,mosser++2012-core-and-evolution-rg-mixed-modes,stello-2013-classifcation-13000-rg-kepler,vrard-2016-period-spacing-rg-2-automated-measures}. We made the diagram in Figure~\ref{fig:dpi1_vs_dnu} for more-evolved subgiants. These stars have more than one avoided crossing, which makes measuring \dpi{1} feasible, although we note that these are lower-order g modes so we are not in the asymptotic regime. For subgiants, stellar models predict \dpi{1} is rather independent of \Dnu{}, especially for higher-mass subgiants, such that they evolve almost vertically \citep{benomar++2013-avoided-crossing-sg-kepler,gai++2017-sb-rb-period-spacing-diagram}. 

Finally, we note that using the two diagrams to infer mass and age provides a method independent of asteroseismic scaling relations. This is discussed in more detail in Paper II.

\section{Mode Linewidths}
\label{sec:lw}

\subsection{Linewidths as a function of frequency}
\label{subsec:lw-nu}
\begin{figure}
	\includegraphics[width=\linewidth]{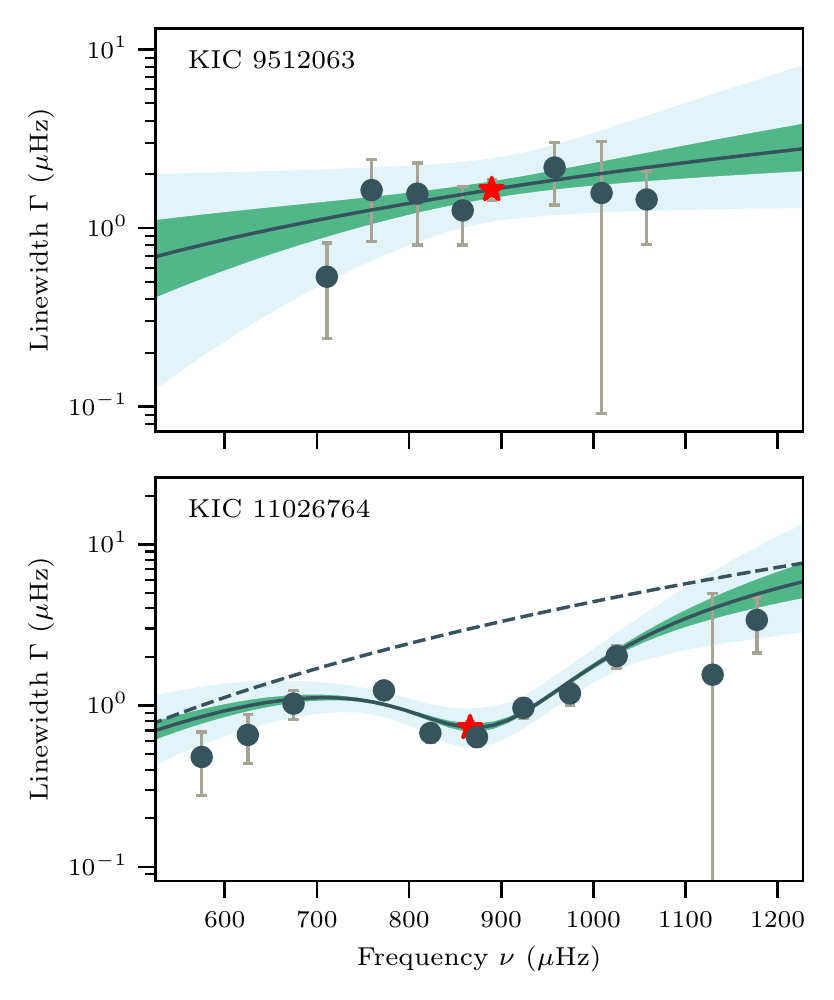}
    \caption{Linewidths as a function of frequencies for KIC 9512063 (top) and Gemma (KIC 11026764, bottom), denoted by the filled circles. The solid lines are the fits of Equation~\ref{eq:gamma-nu}, with KIC 9512063 using the first two terms (the power-law), and Gemma using the full terms. The dashed line is the power-law component in Gemma's fit. The shaded green regions show 1-$\sigma$ and 3-$\sigma$ credible intervals for the fit. The red star symbols around 900 \muhz{} are the estimations of linewidth at \numax{}, obtained from the MCMC samples.}
    \label{fig:lwmax-1}
\end{figure}

\begin{figure}
	\includegraphics[width=\linewidth]{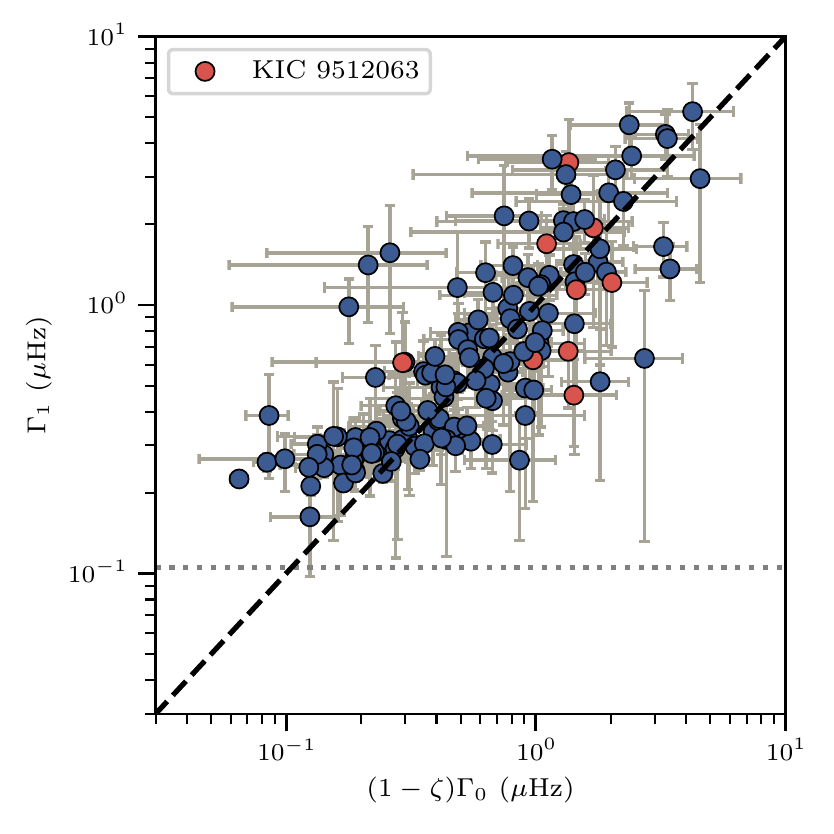}
    \caption{Linewidths of dipole mixed modes vs. those of radial modes modified with the stretch function $\zeta$. The dashed line indicates the 1:1 relation. The horizontal dotted line denotes the nominal frequency resolution, $1/$\tobs{}, of KIC 9512063, which is around $0.1$ \muhz{}; the largest among the sample.}
    \label{fig:lw-dipole}
\end{figure}

\begin{table*} 
\caption{Fit parameters of the $\Gamma$\,-\,$\nu$ relation.} 
\label{tab:lwdip} 
\begin{tabular*}{\textwidth}{@{\extracolsep{\fill}}rrrrrrrrr}
\toprule
      KIC & $\Gamma(\nu_{\rm max})$ &          $\alpha$ & $\Gamma_\alpha$ ($\mu$Hz) & $\Delta\Gamma_\text{dip}$ ($\mu$Hz) & $\nu_\text{dip}$ ($\mu$Hz) & $W_\text{dip}$ ($\mu$Hz) &   FWHM ($\mu$Hz) & $A_\text{dip}$ ($\mu$Hz$^{-1}$) \\
\midrule
  2991448 &         $0.73 \pm 0.22$ &   $1.35 \pm 1.38$ &         $56.96 \pm 55.33$ &                     $0.01 \pm 0.01$ &              $1100 \pm 11$ &            $1477 \pm 93$ &     $311 \pm 72$ &                     $80 \pm 87$ \\
  3852594 &         $4.42 \pm 1.18$ &   $1.35 \pm 0.64$ &           $5.95 \pm 0.40$ &                                ---  &                       ---  &                     ---  &             ---  &                            ---  \\
  5607242 &         $0.43 \pm 0.16$ &   $4.24 \pm 0.71$ &           $2.35 \pm 1.17$ &                     $0.12 \pm 0.04$ &                $698 \pm 7$ &             $898 \pm 76$ &     $191 \pm 60$ &                       $8 \pm 3$ \\
  5689820 &         $0.13 \pm 0.05$ &   $1.71 \pm 1.65$ &         $12.70 \pm 30.15$ &                     $0.01 \pm 0.05$ &               $731 \pm 67$ &          $2534 \pm 1580$ &   $1029 \pm 565$ &                    $69 \pm 233$ \\
  5955122 &         $1.67 \pm 0.41$ &   $3.17 \pm 0.94$ &         $11.83 \pm 28.40$ &                     $0.10 \pm 0.12$ &               $874 \pm 37$ &           $1729 \pm 554$ &    $621 \pm 298$ &                     $10 \pm 12$ \\
  6064910 &         $5.13 \pm 1.33$ &   $0.16 \pm 0.17$ &           $6.59 \pm 0.38$ &                                ---  &                       ---  &                     ---  &             ---  &                            ---  \\
  6442183 &         $0.64 \pm 0.13$ &   $2.93 \pm 0.41$ &           $2.79 \pm 1.44$ &                     $0.20 \pm 0.07$ &              $1166 \pm 12$ &           $1572 \pm 181$ &    $350 \pm 136$ &                       $5 \pm 2$ \\
  6693861 &         $0.59 \pm 0.46$ &   $5.39 \pm 2.75$ &         $42.11 \pm 58.02$ &                     $0.01 \pm 0.03$ &               $838 \pm 39$ &           $1529 \pm 337$ &    $532 \pm 195$ &                    $71 \pm 174$ \\
  7174707 &         $0.42 \pm 0.22$ &   $1.70 \pm 1.41$ &         $18.61 \pm 39.44$ &                     $0.02 \pm 0.07$ &               $806 \pm 27$ &           $1723 \pm 473$ &    $613 \pm 238$ &                    $54 \pm 211$ \\
  7199397 &         $1.15 \pm 0.26$ &   $2.23 \pm 0.34$ &           $1.41 \pm 0.07$ &                                ---  &                       ---  &                     ---  &             ---  &                            ---  \\
  7747078 &         $1.04 \pm 0.16$ &   $3.15 \pm 1.56$ &         $10.96 \pm 34.76$ &                     $0.09 \pm 0.18$ &               $958 \pm 66$ &           $2006 \pm 751$ &    $753 \pm 389$ &                     $11 \pm 20$ \\
  7976303 &         $1.76 \pm 0.26$ &   $2.11 \pm 1.09$ &          $8.63 \pm 34.46$ &                     $0.18 \pm 0.28$ &              $933 \pm 121$ &          $2790 \pm 2583$ &  $1155 \pm 1027$ &                       $6 \pm 9$ \\
  8524425 &         $0.60 \pm 0.65$ &   $4.64 \pm 1.13$ &           $2.97 \pm 3.81$ &                     $0.21 \pm 0.13$ &              $1096 \pm 35$ &           $1603 \pm 284$ &    $447 \pm 199$ &                       $5 \pm 3$ \\
  8702606 &         $0.31 \pm 0.09$ &   $3.14 \pm 0.95$ &         $12.17 \pm 32.95$ &                     $0.04 \pm 0.10$ &               $660 \pm 26$ &           $1547 \pm 512$ &    $572 \pm 239$ &                     $23 \pm 55$ \\
  9512063 &         $1.21 \pm 0.63$ &   $1.75 \pm 0.92$ &           $1.48 \pm 0.18$ &                                ---  &                       ---  &                     ---  &             ---  &                            ---  \\
 10018963 &         $2.37 \pm 0.42$ &   $2.60 \pm 0.22$ &           $2.37 \pm 0.09$ &                                ---  &                       ---  &                     ---  &             ---  &                            ---  \\
 10147635 &         $2.79 \pm 0.73$ &   $3.71 \pm 0.43$ &           $2.43 \pm 0.16$ &                                ---  &                       ---  &                     ---  &             ---  &                            ---  \\
 10273246 &         $1.68 \pm 0.46$ &   $5.07 \pm 1.08$ &          $8.51 \pm 21.03$ &                     $0.23 \pm 0.31$ &             $1359 \pm 695$ &          $5690 \pm 2432$ &  $2919 \pm 1719$ &                       $4 \pm 6$ \\
 10593351 &         $2.03 \pm 0.64$ &   $3.04 \pm 0.44$ &           $2.29 \pm 0.17$ &                                ---  &                       ---  &                     ---  &             ---  &                            ---  \\
 10873176 &         $3.26 \pm 1.59$ &   $3.67 \pm 2.10$ &           $3.42 \pm 0.76$ &                                ---  &                       ---  &                     ---  &             ---  &                            ---  \\
 10920273 &         $0.57 \pm 0.25$ &   $8.48 \pm 3.66$ &         $21.88 \pm 44.10$ &                     $0.03 \pm 0.07$ &              $1082 \pm 35$ &           $1577 \pm 443$ &    $459 \pm 311$ &                     $33 \pm 81$ \\
 10972873 &         $0.67 \pm 0.14$ &   $3.69 \pm 1.40$ &         $13.36 \pm 38.59$ &                     $0.05 \pm 0.14$ &              $1062 \pm 52$ &           $2204 \pm 932$ &    $823 \pm 483$ &                     $18 \pm 46$ \\
 11026764 &         $0.73 \pm 0.17$ &   $2.17 \pm 0.43$ &           $1.99 \pm 0.36$ &                     $0.31 \pm 0.06$ &               $882 \pm 10$ &            $1065 \pm 51$ &     $183 \pm 43$ &                       $3 \pm 1$ \\
 11137075 &         $0.53 \pm 0.22$ &  $10.02 \pm 1.98$ &         $49.24 \pm 55.66$ &                     $0.01 \pm 0.01$ &              $1224 \pm 25$ &           $1912 \pm 173$ &    $601 \pm 115$ &                   $110 \pm 173$ \\
 11193681 &         $0.86 \pm 0.22$ &   $3.06 \pm 0.37$ &           $1.20 \pm 0.06$ &                                ---  &                       ---  &                     ---  &             ---  &                            ---  \\
 11395018 &         $0.67 \pm 0.15$ &   $1.72 \pm 1.56$ &         $35.73 \pm 51.69$ &                     $0.02 \pm 0.03$ &               $822 \pm 21$ &           $1497 \pm 200$ &    $469 \pm 115$ &                     $45 \pm 70$ \\
 11414712 &         $0.69 \pm 0.16$ &   $2.66 \pm 0.26$ &           $0.94 \pm 0.04$ &                                ---  &                       ---  &                     ---  &             ---  &                            ---  \\
 11771760 &         $0.63 \pm 0.17$ &   $0.79 \pm 0.52$ &           $1.05 \pm 0.07$ &                                ---  &                       ---  &                     ---  &             ---  &                            ---  \\
 12508433 &         $0.27 \pm 0.06$ &   $3.92 \pm 1.66$ &         $26.15 \pm 44.41$ &                     $0.01 \pm 0.03$ &               $796 \pm 56$ &           $2294 \pm 556$ &    $890 \pm 230$ &                    $73 \pm 154$ \\
\bottomrule
\end{tabular*}
\end{table*}

The modes of solar-like oscillations are damped by convection. For these observations, the modes have lifetimes much shorter than the length of time series, which results in a broadened Lorentzian shape in the power spectrum. The modes are hence ``resolved''. The linewidth (full-width-at-half-maximum) of that shape can be translated to the lifetime through $\tau = (\pi\Gamma)^{-1}$, or damping rate $\eta=\pi\Gamma$. This is important to constrain the physics of the super-adiabatic layers near the surface, including the mechanism of how the kinetic energy of oscillations is dissipated into turbulent convection. 
The answer requires a detailed treatment of convection, but it is difficult to formulate with current convection theories
\citep{gough-1980-theoretical-remarks-on-stellar-oscillations, balmforth-1992-pulsation-stability-1-mode-thermodynamics, gabriel-1996-stellar-oscillation-unno-convection, houdek++1999-amplitudes-solarlike,  grigahcene++2005-mlt-convection-theory-oscillation, xiong++2015-turbulence-stabiltiy-basic-equation}. However, the theories are able to reproduce the frequency-dependence of the linewidths seen in the observations, provided the free parameters in the modelling are carefully calibrated, in red giants \citep{aarslev++2018-linewidth-ngc6819} and in MS stars \citep{Houdek++2019-damping-rate-lagacy-kepler}. 3D hydrodynamical convection simulations may offer a promising path, by eliminating the degrees of freedom in the theory and provide a numerical solution \citep[e.g.][]{belkacem++2019-3d-rhd,zhou++2019-3d-amplitude-solar-oscillation}. 


\citet{appourchaux++2014-lw-height-ms-kepler} first parametrised the $\Gamma$\,--\,$\nu$ relation in a given star and \citet{lund++2017-legacy-kepler-1} improved the fitting details. Here, we followed their procedure to fit the linewidths of radial modes, as follows:
\beq\label{eq:gamma-nu} \ln\Gamma(\nu; \theta) = \alpha\ln(\nu/\nu_{\rm max})+\ln\Gamma_\alpha + \frac{\ln\Delta\Gamma_{\rm dip}}{1+\left[ \frac{2\ln(\nu/\nu_{\rm dip})}{\ln(W_{\rm dip}/\nu_{\rm max})}\right]^2 },\eeq
where the free parameters are $\theta=(\alpha, \Gamma_\alpha, \Delta\Gamma_{\rm dip}, \nu_{\rm dip}, W_{\rm dip})$. This $\Gamma$\,--\,$\nu$ relation follows a power-law but saturates near \numax{} to form a plateau (or a dip) centred on $\nu_{\rm dip}$, with width $W_{\rm dip}$ and height $\Delta\Gamma_{\rm dip}$. The dip could originate from a resonance of the thermal timescale in the superadiabatic layer and mode frequency \citep{balmforth-1992-pulsation-stability-1-mode-thermodynamics,belkacem++2011-physics-under-numax-nuc}. 
To fit the parameters, we optimised the likelihood function:
\beq \ln L \propto \sum_{i} \left[ \ln\Gamma_i - \ln\Gamma(\nu_i;\theta) \right]^2 / \sigma_{\ln\Gamma_i}^2.\eeq
The modes used in this fit were radial modes satisfying Bayes factors $\ln K>1$. The priors of $\Delta\Gamma_{\rm dip}$ and $W_{\rm dip}$ deserve a further mention. As suggested by \citet{lund++2017-legacy-kepler-1}, the values of $\Delta\Gamma_{\rm dip}$ were bound to be between 0 and 1 to reduce correlations between the parameters. $W_{\rm dip}$ can have two solutions, one larger than \numax{}, and one smaller. We kept the convention from \citet{lund++2017-legacy-kepler-1} to use the larger one. We estimated the FWHM (full-width-at-half-maximum) of the dip through FWHM$=\nu_{\rm dip}\left|\sqrt{W_{\rm dip}/\nu_{\rm max}} - \sqrt{\nu_{\rm max}/W_{\rm dip}}\right|$ \citep{appourchaux++2016-lw-height-ms-kepler-corrigendum}, and the amplitude of the dip as $A_{\rm dip}=\exp{|\ln\Delta\Gamma_{\rm dip}|}$ \citep{lund++2017-legacy-kepler-1}.
Figure~\ref{fig:lwmax-1} shows the fit for Gemma (KIC 11026764), which shows a dip around \numax{}. Note that some stars do not present the evidence of a dip. Their linewidths increase monotonically with frequencies. We only fitted those stars with the first two terms of Equation~\ref{eq:gamma-nu}. Figure~\ref{fig:lwmax-1} also shows an example, KIC 9512063, where the large random errors obscure the presence of either a dip or a plateau.
Table~\ref{tab:lwdip} lists the fitted parameters for all stars.

The dipole modes have a more complicated $\Gamma$\,--\,$\nu$ relation. The dipole modes with more g mode characteristics have larger inertia and are less affected by damping from the surface than are radial modes, hence resulting in smaller linewidths. \citet{benomar++2014-kepler-mode-inertia} suggested using linewidths to measure mode inertias, which could also be obtained from the asymptotic formalism of mixed modes \citep{goupil++2013-am-transport-2-rot-splitting-rg,shibahashi-1979-modal-analysis-nonradial-oscillation-asymptotic-method}. \citet{mosser++2018-period-spacing-4-complete-description-mixed-mode-patterns} introduced a stretch function $\zeta$, which is the degree of mode trapping characterised by the ratio of mode inertia inside the g-mode cavity to that throughout the star. The linewidths of mixed modes, $\Gamma_1$, relate to those of radial modes, $\Gamma_0$, as follows \citep{mosser++2011-rg-oscillation-pattern-automated-corot, mosser++2015-period-spacing-1-rotation-core-structure-discontinuities, vrard-2016-period-spacing-rg-2-automated-measures,hekker+2016-giant-seismology-review,mosser++2018-period-spacing-4-complete-description-mixed-mode-patterns}:
\beq \Gamma_1 = \Gamma_0(1-\zeta),\eeq
where 
\beq \zeta(\nu) = \left[ 1+ \frac{q}{N(\nu)}\frac{1}{q^2 \cos^2\theta_p(\nu) + \sin^2\theta_p(\nu)} \right]^{-1},\eeq
\beq N(\nu) = \frac{\Delta\nu}{\nu^2\Delta\Pi_1}, \eeq
and
\beq \theta_p(\nu) = \pi \left\{ \frac{\nu}{\Delta\nu}  - \left[ n_p+\frac{l}{2}+\epsilon_p-  d_{0l}+ 
\frac{\alpha_p}{2} \left(n_p-\frac{\nu_{\rm max}}{\Delta\nu} \right)^2 \right]  \right\}, \eeq
Using the above definition of $\theta_p$, together with Equation~\ref{eq:mmode} and~\ref{eq:thetag}, we fitted the asympotic parameters $\{\epsilon_p, \alpha_p, d_{01}, \Delta\nu, \Delta\Pi_1, \epsilon_g\}$ to the dipole mode frequencies estimated from the peakbagging. The value of $\zeta$ was then straightforwardly obtained. In Figure~\ref{fig:lw-dipole}, we compare $\Gamma_1$ with $\Gamma_0(1-\zeta)$ and find good agreement. That is, the widths of the dipole modes are smaller than those of the radial modes, as expected. This suggests that the asymptotic theory agrees well with the observation. 

\subsection{Scaling relations for linewidths}

\begin{table} 
\caption{Fit parameters of the $\Gamma(\nu_{\rm max})$ scaling relation.} 
\label{tab:lwscaling} 
\begin{tabular*}{\columnwidth}{@{\extracolsep{\fill}}lrrr}
\toprule
     Phase &       $\Gamma_0$ &             $c_1$ &             $c_2$ \\
\midrule
        MS &  $1.10 \pm 0.01$ &  $13.41 \pm 0.10$ &   $0.05 \pm 0.02$ \\
 Subgiants &  $0.50 \pm 0.04$ &  $13.71 \pm 0.19$ &  $-0.65 \pm 0.06$ \\
       RGB &  $0.25 \pm 0.00$ &   $2.78 \pm 0.09$ &   $0.09 \pm 0.00$ \\
       HeB &  $0.47 \pm 0.02$ &   $1.70 \pm 0.13$ &   $0.19 \pm 0.01$ \\
\bottomrule
\end{tabular*}
\end{table}




\begin{figure}
	\includegraphics[width=\linewidth]{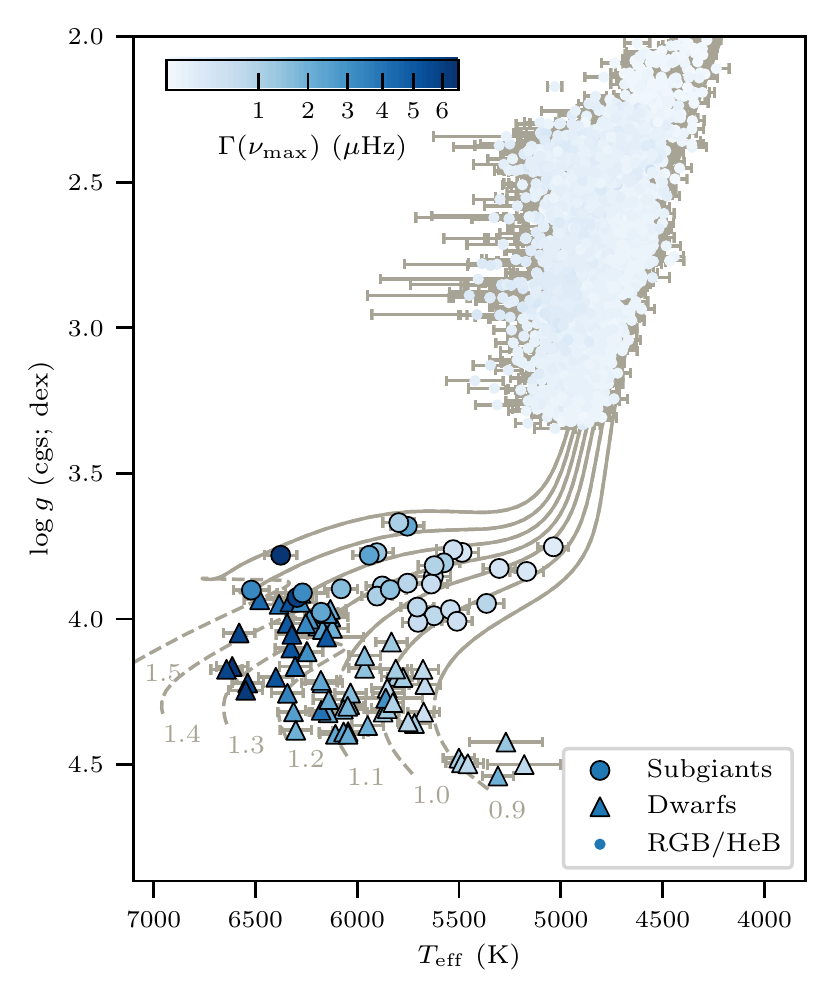}
    \caption{Kiel diagram with colour-coded \lwmax{}. The measurements of dwarfs and red giants are adopted from \citet{lund++2017-legacy-kepler-1} and \citet{vrard++2018-rgb-radial-mode-amplitude-lifetime}, respectively. The theoretical evolutionary tracks \citep{Choi++2016-mist-1-solar-scaled-models} before and after the MS turnoff are shown in the dashed and solid lines, respectively, with each track labelled with mass in solar units.}
    \label{fig:lwmax_hrd}
\end{figure}

Scaling relations for linewidths were studied only recently, focusing on the linewidths around \numax{}, denoted by $\Gamma(\nu_{\rm max})$. \citet{chaplin++2009-lifetime} used non-adiabatic pulsation computations and ground-based observations to study the linewidths across different stages of evolution. They showed that the linewidths scale with fundamental stellar parameters, primarily \teff{}. 
This dependence on surface properties is expected because damping mainly happens in the super-adiabatic regions near the surface \citep{goldreich+1991-thermal-and-mechanical-damping-solar-pmodes}. The studies of \corot{} and \kepler{} targets also suggested that \lwmax{} correlates mainly with \teff{}, but only weakly depends on \logg{} \citep{baudin++2011-amplitude-lw-corot-rg-ms,appourchaux++2014-lw-height-ms-kepler,vrard++2018-rgb-radial-mode-amplitude-lifetime}.

To estimate \lwmax{}, we used the MCMC samples from the fit of $\Gamma$\,--\,$\nu$ relation (Section~\ref{subsec:lw-nu}) to draw the probability distribution of $\Gamma$ at \numax{}. This method assumes that $\Gamma$\,--\,$\nu$ follows the power-law model (or the power-law with a dip). Figure~\ref{fig:lwmax-1} shows an estimation for Gemma (KIC 11026764), marked by a red star around 700 \muhz{}.

In Figure~\ref{fig:lwmax_hrd}, we show \lwmax{} for our subgiant sample, together with MS stars from \citet{lund++2017-legacy-kepler-1}, and RGB and HeB stars from \citet{vrard++2018-rgb-radial-mode-amplitude-lifetime}. Note that \citet{vrard++2018-rgb-radial-mode-amplitude-lifetime} calculated \lwmax{} for the giants differently to how it was done for subgiants and MS stars by averaging linewidths using three modes near \numax{}. Figure~\ref{fig:lwmax_hrd} shows that \lwmax{} mainly depends on \teff{}, and weakly correlates with \logg{}. To parametrise this relation, we fitted \lwmax{} using
\beq\label{eq:lw-power-law-teff-logg} \log(\Gamma/\Gamma_0) = c_1 \log(T_{\rm eff}/5777\ {\rm K}) + c_2 \log (g/274\ {\rm m\cdot s^{-1}}).\eeq
We optimised the likelihood function:
\beq \ln L \propto \left[ \ln \Gamma_i - \ln \Gamma(T_i,g_i;\theta) \right]^2 / (2\sigma_{\ln\Gamma_i}^2).\eeq 
Some previous works \citep{belkacem++2012-damping-rates-hr-diagram,samadi++2015-stellar-oscillation-2-nonadiabatic-case-review} pointed out that \lwmax{} depends on \teff{} and \logg{} differently in each evolutionary stage, which motivates us to fit them separately as well. 

The fitted parameters are shown in Table~\ref{tab:lwscaling}. From the table, we conclude that \lwmax{} primarily depends on \teff{} but with significantly different power-law indices ($c_1$) for red giants and non-red giants. This could be attributed to different dominant damping mechanisms, as some works suggested \citep[e.g.][]{baudin++2011-amplitude-lw-corot-rg-ms}. However it could also be an artefact while selecting modes to compute \lwmax{} for red giants \citep{belkacem++2012-damping-rates-hr-diagram}. The power-law index for \logg{} ($c_2$) in subgiants has a reversed sign compared to MS stars and red giants. This effect could be slightly spotted in Figure~\ref{fig:lwmax_hrd}. 

\section{Mode Amplitudes}
\label{sec:amp}

\begin{figure}
	\includegraphics[width=\linewidth]{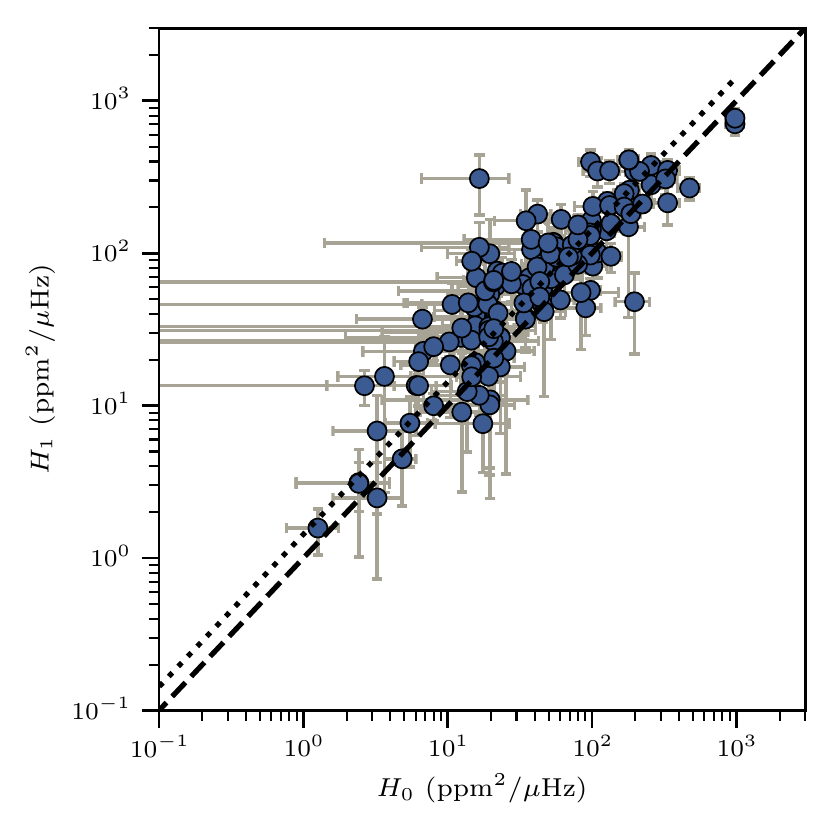}
    \caption{Height of dipole mixed modes $H_1$ vs. those of radial modes $H_0$. The dashed line indicate the 1:1 relation, and the dotted line shows 1.5:1 relation. The factor of 1.5 shows the median value of our observed $H_1/H_0$, which stems from a geometric effect.}
    \label{fig:amp-dipole}
\end{figure}

\begin{figure}
 	\includegraphics[width=\linewidth]{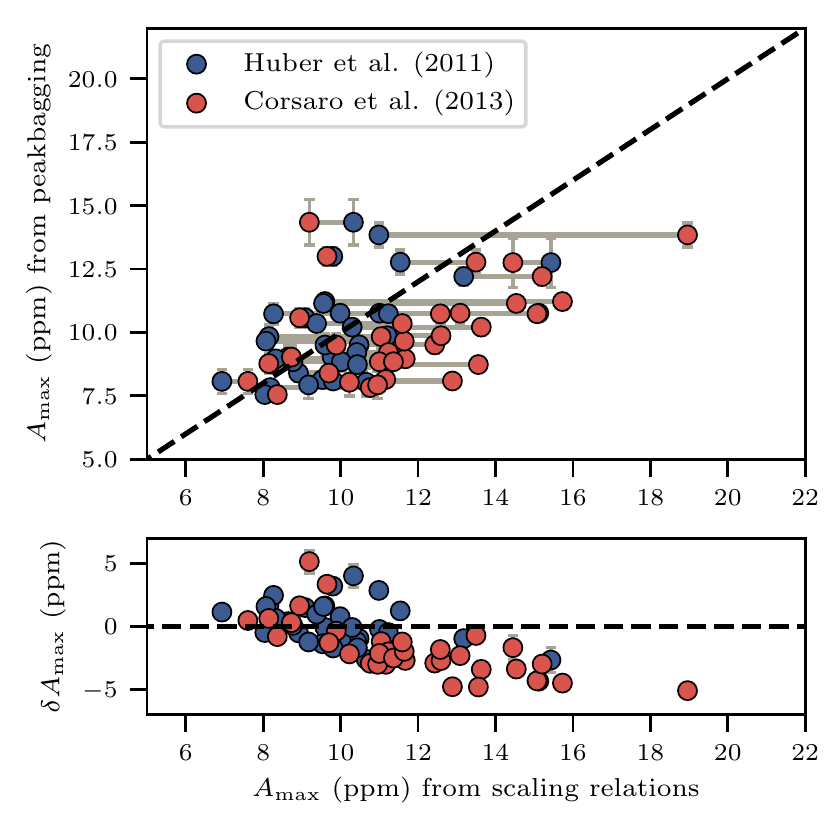}
     \caption{Top: \amax{} estimated using the radial mode amplitudes vs. \amax{} from the scaling relations. The dashed line indicates the 1:1 relation. Bottom: the absolute differences in the top panel.}
     \label{fig:amax_scaling}
\end{figure}

\begin{figure}
	\includegraphics[width=\linewidth]{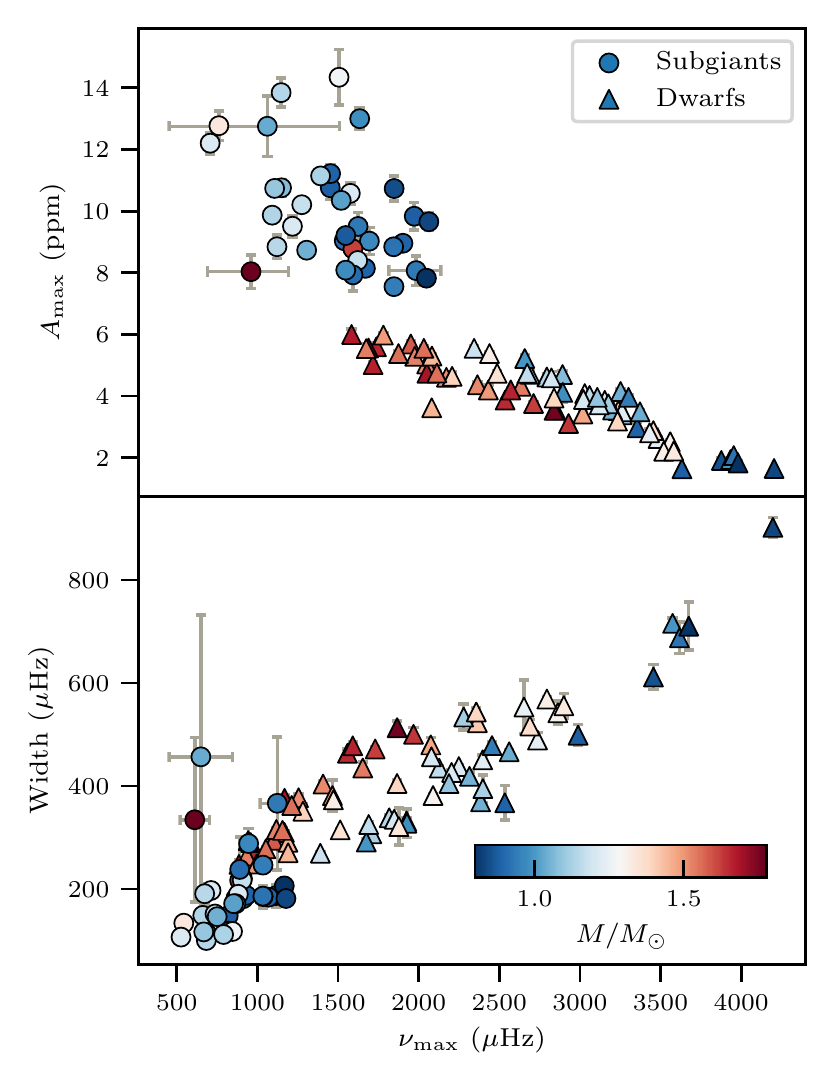}
    \caption{\amax{} and width of the Gaussian envelope against \numax{}. The colour denotes stellar mass.}
    \label{fig:amp-mass-effect}
\end{figure}

\subsection{Amplitudes as a function of frequency}
\label{subsec:amp-nu}

The amplitudes of radial modes follow a roughly Gaussian distribution centred on \numax{}, as we analysed in Section~\ref{sec:freq}. The dipole mixed modes have an added dependence on mode inertia, but are still predictable using the stretch function: $A_1^2 = (1-\zeta)A_0^2$, with $A_0$ being the radial mode amplitudes \citep{benomar++2014-kepler-mode-inertia,belkacem++2015-am-rg-2-spin-down-obs,mosser++2018-period-spacing-4-complete-description-mixed-mode-patterns}. Considering $\Gamma_1 = (1-\zeta)\Gamma_0$, we obtained $H_1=H_0$, suggesting similar mode heights. In Figure~\ref{fig:amp-dipole}, we show $H_1$ against $H_0$. The points deviate from the 1:1 relation because the height of dipole modes are influenced by a geometric effect. Dividing $H_1$ by $H_0$, we get the visibility factor $V_1^2$ with a median value equal to 1.5, similar to the one calculated using stellar atmosphere models \citep{ballot++2011-visibility-kepler-seismology}, which is also around 1.5.

\subsection{Scaling relations for amplitudes}
\citet{kjeldsen+1995-scaling-relations} first proposed a scaling relation for the amplitudes of solar-like oscillations to scale from the Sun to other stars, relating both radial velocity and photometric amplitude to fundamental parameters $L$, $M$ and \teff{}. It received numerous application, e.g. for optimising target selections \citep{chaplin++2011-detectability-solartype-kepler,schofield++2019-tess-atl}, and understanding how modes are excited with different treatment for convection and excitation processes \citep[e.g.][]{goldreich+1977-2-stochastic-excitation-turbulent-convection,balmforth-1992-pulsational-stability-3-turbulence,samadi+2001-excitation-stellar-pmodes-1-theory,chaplin++2005-model-prediction-solar-p-mode-psd,zhou++2019-3d-amplitude-solar-oscillation}. As we mentioned in Section~\ref{sec:freq}, while estimating \numax{} we also measured the maximum photometric amplitude expressed by \amax{} and the width of the Gaussian-like envelopes. To compare \amax{} with scaling relations, we used two results from \citet{huber++2011-test-scaling-relation-ms-rgb-kepler}:
\beq 
\frac{A_\text{max}}{A_{\text{bol},\astrosun}}= c_\text{K} \left(\frac{L}{L_{\astrosun}}\right)^{0.838}\left(\frac{M}{M_{\astrosun}}\right)^{-1.32}\left(\frac{T_\text{eff}}{T_{\text{eff},\astrosun}}\right)^{-1}, 
\eeq
and \citet{corsaro++2013-bayesian-scaling-relations-amplitudes-kepler}:
\beq 
\frac{A_\text{max}}{A_{\text{bol},\astrosun}}=1.38 c_\text{K} \left(\frac{\nu_\text{max}}{\nu_{\text{max},\astrosun}}\right)^{-2.314}\left(\frac{\Delta\nu}{\Delta\nu_{\astrosun}}\right)^{2.088}\left(\frac{T_\text{eff}}{T_{\text{eff},\astrosun}}\right)^{0.365},
\eeq
where $A_{\text{bol},\astrosun}=3.6\pm0.11$ ppm and $c_\text{K}=(T_\text{eff}/5934\ \text{K})^{0.8}$\citep{huber++2011-test-scaling-relation-ms-rgb-kepler}. 
There has been some discussions concerning the exponent of each term \citep[e.g.][]{houdek++1999-amplitudes-solarlike,houdek-2006-excitation-damping-solartype,samadi++2007-excitation-solar-like-oscillation}, and the above two papers set the exponents free and calibrated with observations \citep[see also e.g.][]{stello++2011-amplitude-clusters}. Figure~\ref{fig:amax_scaling} shows the comparison. In the regime of subgiants, \citet{huber++2011-test-scaling-relation-ms-rgb-kepler}'s relation seems to fit better, with a correlation coefficient of 0.55 against 0.49 from \citet{corsaro++2013-bayesian-scaling-relations-amplitudes-kepler}. However, both relations overestimate the amplitudes, by factors of 9\% and 24\% respectively. The scatter of the data points could be caused by activity \citep{chaplin++2011-activity-solartype-oscillation-kepler} or an unaccounted metallicity effect \citep{yuj++2018-asteroseismology-16000-rg}. We also verified that the suppression effect of non-radial modes seen in red giants \citep{mosser++2012-amplitude-rg,stello-2016-supression-rg-kepler} does not affect the subgiants in our sample \citep{garcia++2014-rotation-magnetism-kepler-age-rotation-relations,fuller-2015-asteroseismology-strong-Bfield-rg-science}.

In Figure~\ref{fig:amp-mass-effect} we plot both \amax{} and the width against \numax{}. The values of \amax{} increase and the widths decrease with decreasing \numax{}, indicating that the Gaussian envelope becomes higher and narrower as the star evolves. The points in Figure~\ref{fig:amp-mass-effect} are also colour-coded with mass, and we identify a weak correlation: at a given \numax{}, \amax{} is smaller in higher-mass stars, with the width being larger. This phenomenon is also seen in the \kepler{} red giant sample \citep[e.g.][]{yuj++2018-asteroseismology-16000-rg}. We also analysed the effect of metallicity on mode amplitude but due to the small size of our sample we were not able to draw any conclusions.

\section{Conclusions}
\label{sec:conc}

In this paper, we presented oscillation frequencies, linewidths and amplitudes for 36 subgiants observed by the 4 year \emph{Kepler} mission. We derived those parameters with uncertainties from MCMC fitting, using an open-source software package \textsc{solarlikepeakbagging}. Significance tests and visual inspections were applied to ensure a robust identification. Our main results are summarised as follows:
\begin{enumerate}
    \item With the long baseline of \kepler{} observations, the median value for the frequency uncertainties of the subgiants is 0.180 \muhz{}. For modes with S/N=3, the typical uncertainty is 0.1 \muhz{}, providing strong constraint on stellar models with these data. 
    \item The asymptotic parameters \epsp{}, \Dnu{}, \numax{} and \dnu{0l} were derived using the mode parameters. We identified a \teff{} and \logg{} dependence for \epsp{}. We also revisited the C\,--\,D diagram and the \Dnu{}\,--\,\numax{} diagram with a focus on subgiants. The subgiants deviate slightly from the general trend, but could be explained using simple scaling arguments or stellar models.
    \item We presented two p\,--\,g diagrams, the \nug{1}\,--\,\Dnu{} diagram for less-evolved subgiants and the \dpi{1}\,--\,\Dnu{} diagram for more-evolved subgiants. Both diagrams were populated with \kepler{} observations and can be used as a first simple estimation of stellar mass.
    \item The linewidths of radial modes were analysed as functions of frequencies, and the linewidths of dipolar mixed modes agree well with asymptotic predictions. We verified the dependence of \lwmax{} on \teff{} and also observed a weak \logg{} effect.
    \item The amplitudes of dipolar mixed modes also agree with asymptotic predictions. The mass dependence of \amax{} and the width is present both in subgiants and MS stars.
\end{enumerate}
The unprecedented quality of asteroseismic data presented by this sample is valuable for the study of stellar physics. The mode frequencies are further used as modelling input in Paper II. The linewidths and amplitudes derived in this work can be used to study mode excitation and damping. In the observation data, one missing piece that remains unexplored in this work is the rotational splitting, which provides a golden opportunity to study the angular momentum transport in subgiants. We will continue to study this topic in this series of papers.

\section*{Acknowledgements}
We thank Mikkel Lund and J\o{}rgen Christensen-Dalsgaard for very helpful discussions. We also appreciate efforts from the referee to improve the quality of this paper, especially during a pandemic outbreak. The \kepler{} Discovery mission is funded by NASA's Science Mission Directorate. We acknowledge the Joint Research Fund in Astronomy (U1631236) under cooperative agreement between the National Natural Science Foundation of China (NSFC) and Chinese Academy of Sciences (CAS), (NSFC 11273007 and 10933002), and the Fundamental Research Funds for the Central Universities. We also acknowledge funding from the Australian Research Council.

\bibliographystyle{mnras}
\bibliography{ref/myastrobib,ref/pkbg_kepler_subgiant}

\appendix

\section{Power spectra of 36 subgiants. }
Left: \'{e}chelle diagrams. The blue circles ($l=0$), red upward triangles ($l=1$), green squares ($l=2$) and purple downward triangles ($l=3$) mark the extracted frequencies. Right: power spectra. The fitted power spectra (black) are overlaid on the original power spectra (light grey) and the 1.0 $\mu$Hz smoothed spectra (grey).

\begin{figure*} 
\includegraphics[width=0.94\linewidth]{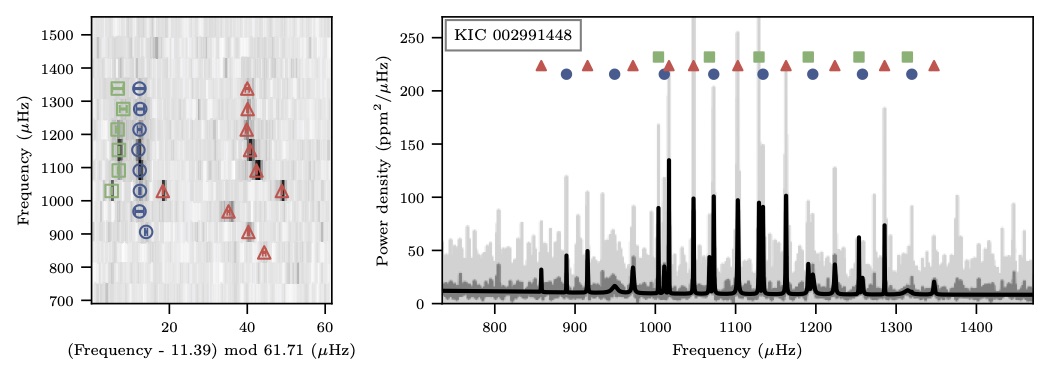} 
\includegraphics[width=0.94\linewidth]{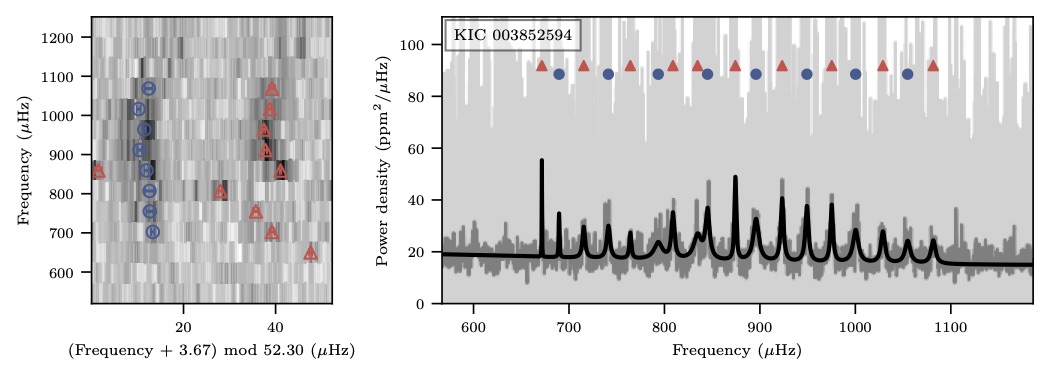} 
\includegraphics[width=0.94\linewidth]{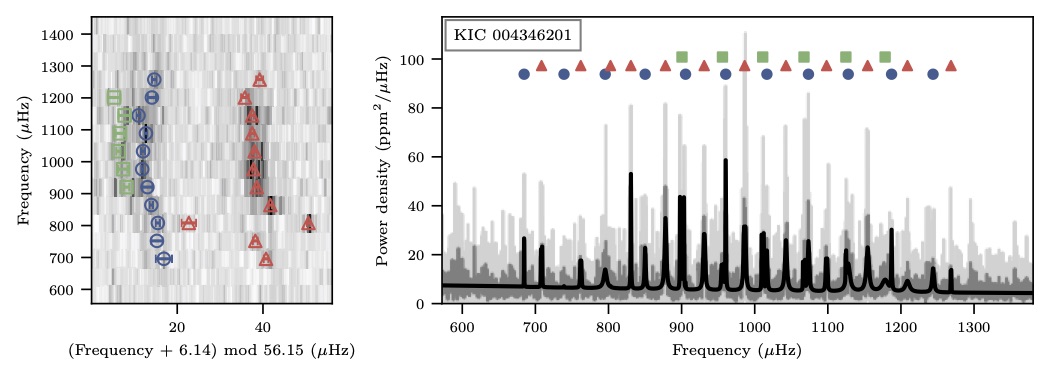} 
\includegraphics[width=0.94\linewidth]{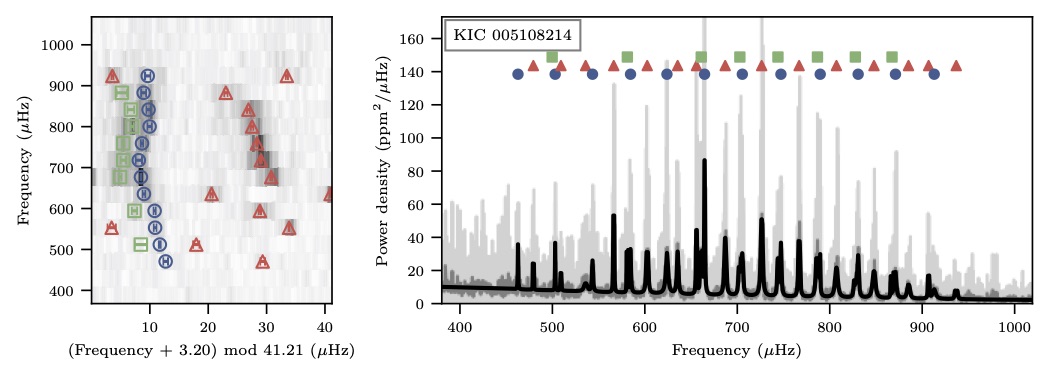} 
\end{figure*} 
\clearpage 
\begin{figure*} 
\includegraphics[width=0.94\linewidth]{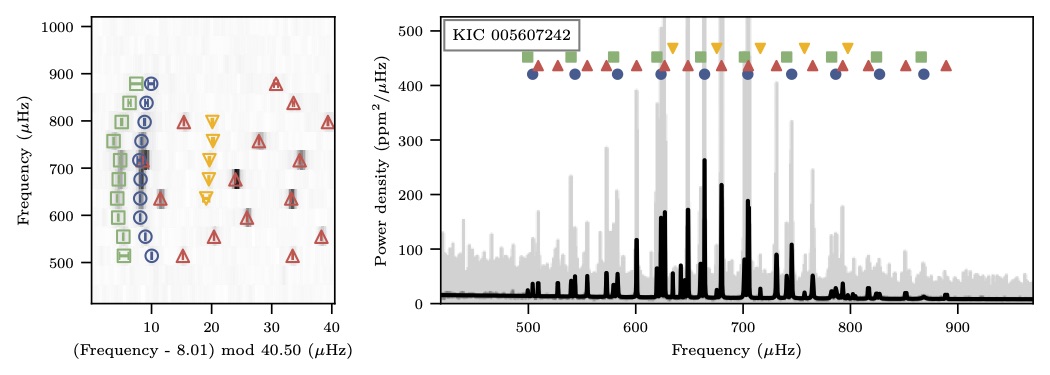} 
\includegraphics[width=0.94\linewidth]{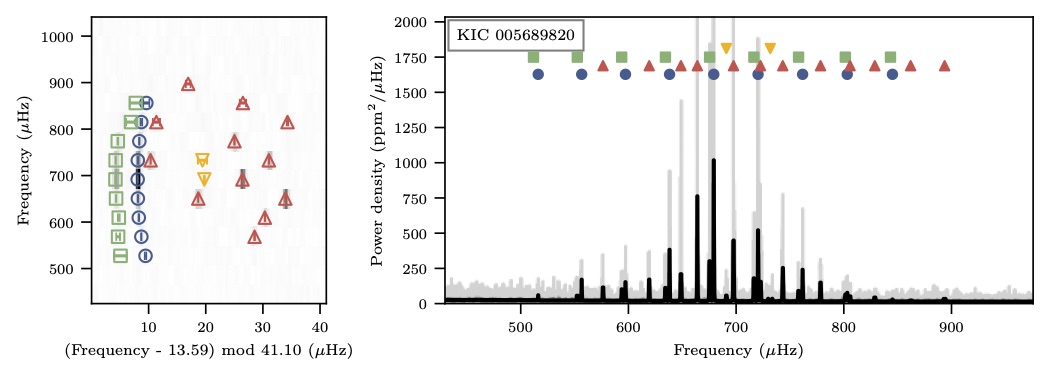} 
\includegraphics[width=0.94\linewidth]{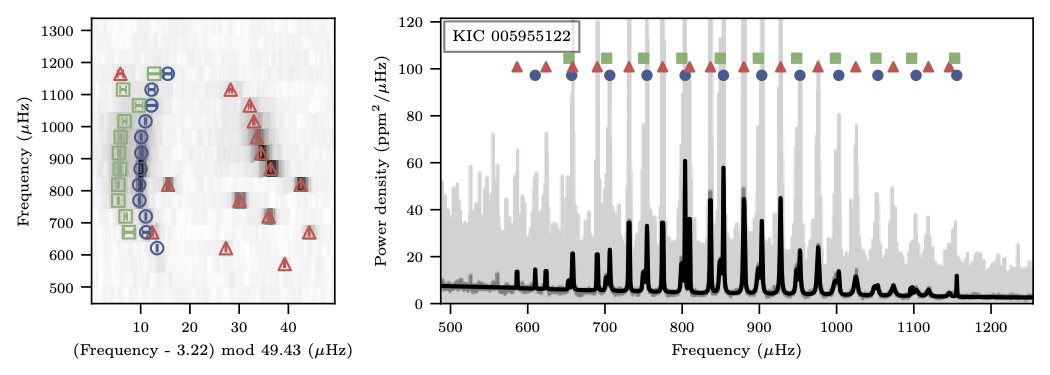} 
\includegraphics[width=0.94\linewidth]{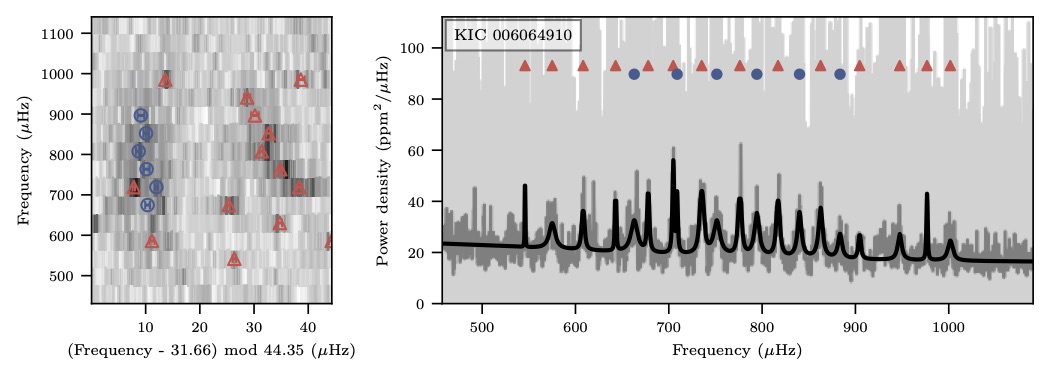} 
\end{figure*} 
\clearpage 
\begin{figure*} 
\includegraphics[width=0.94\linewidth]{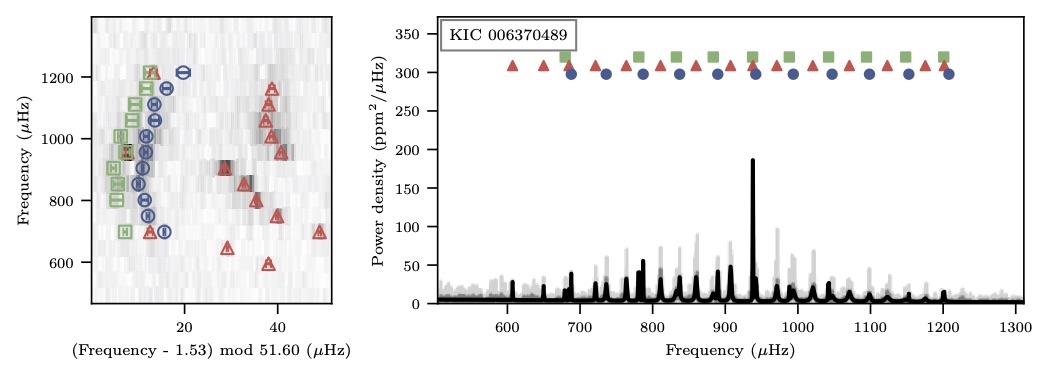} 
\includegraphics[width=0.94\linewidth]{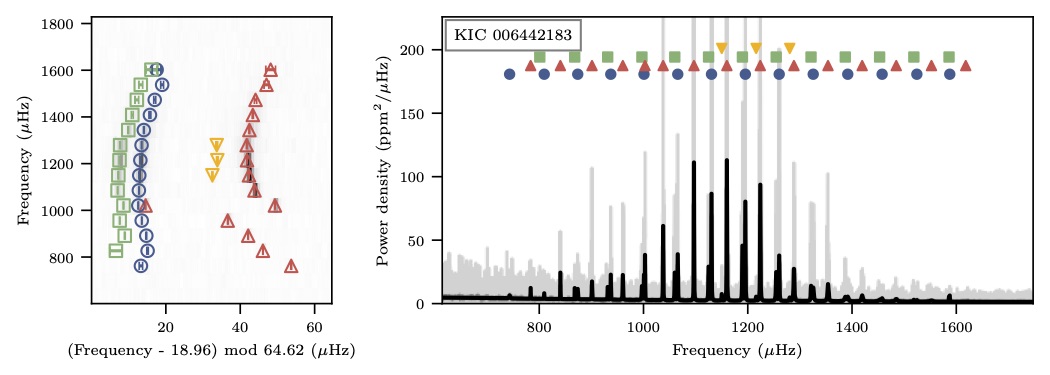} 
\includegraphics[width=0.94\linewidth]{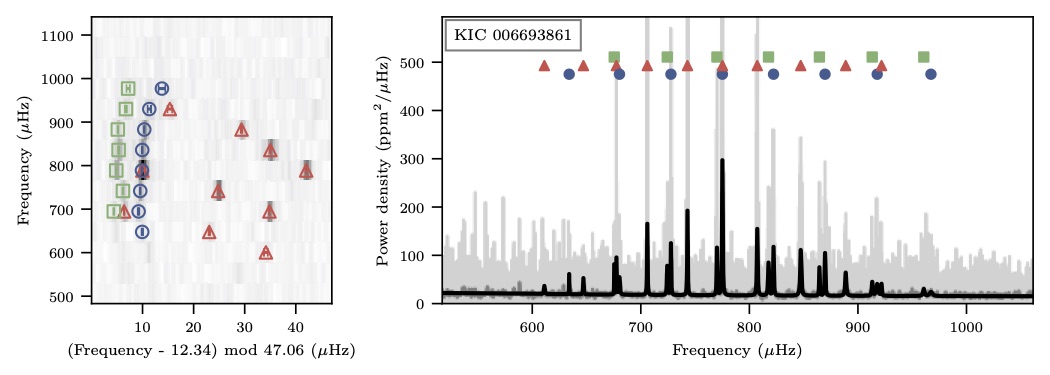} 
\includegraphics[width=0.94\linewidth]{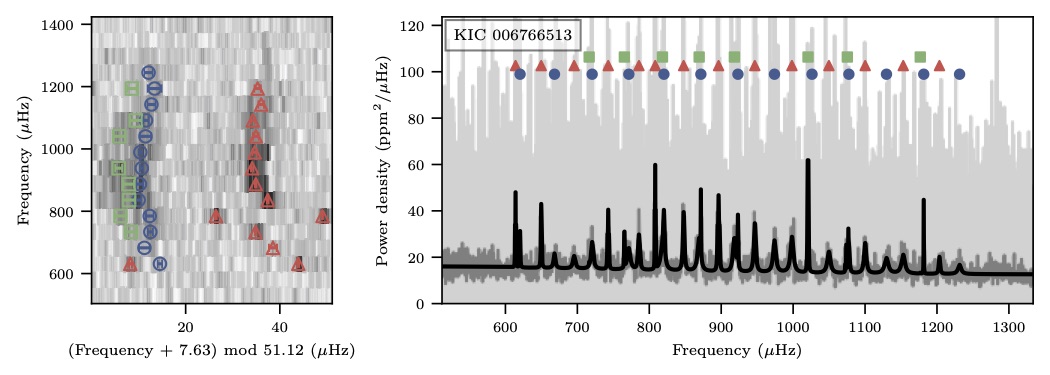} 
\end{figure*} 
\clearpage 
\begin{figure*} 
\includegraphics[width=0.94\linewidth]{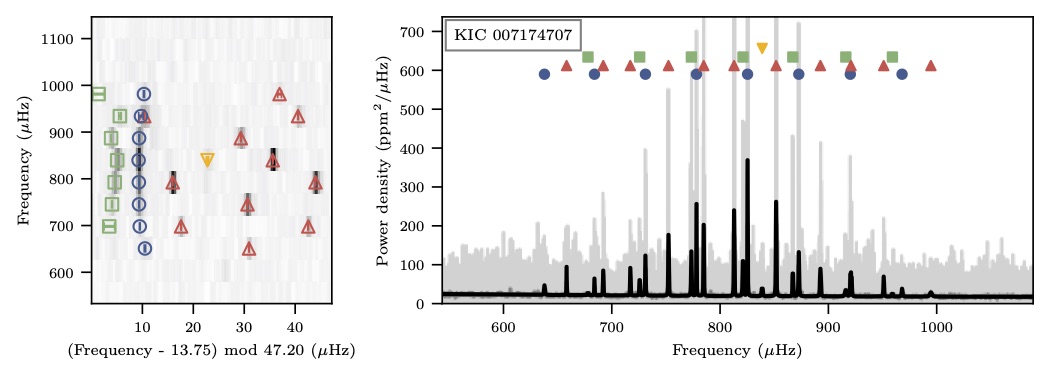} 
\includegraphics[width=0.94\linewidth]{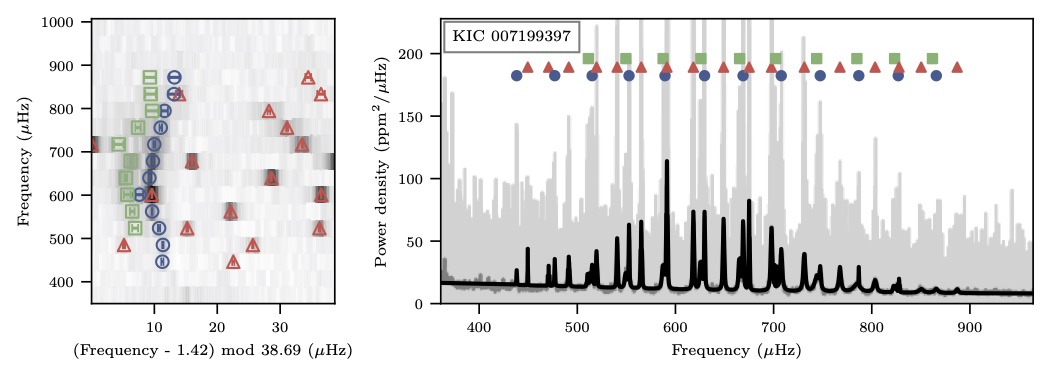} 
\includegraphics[width=0.94\linewidth]{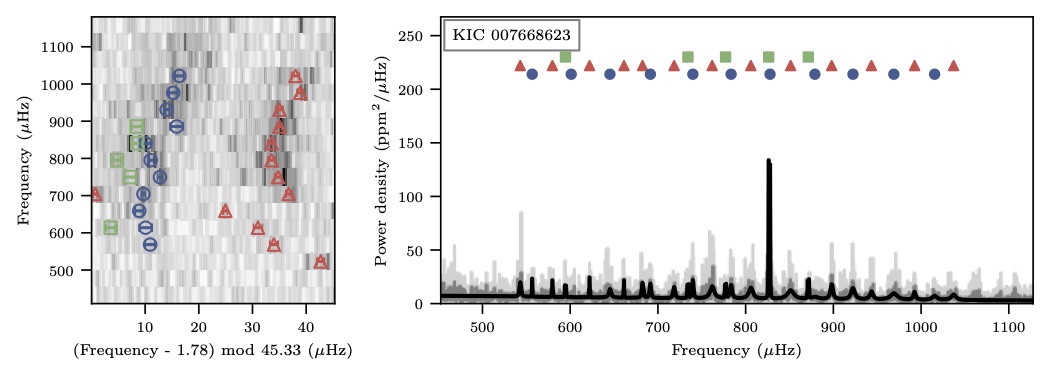} 
\includegraphics[width=0.94\linewidth]{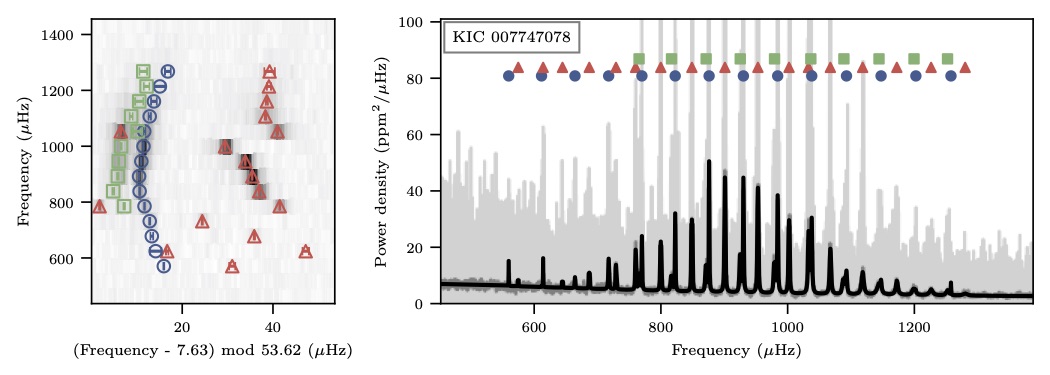} 
\end{figure*} 
\clearpage 
\begin{figure*} 
\includegraphics[width=0.94\linewidth]{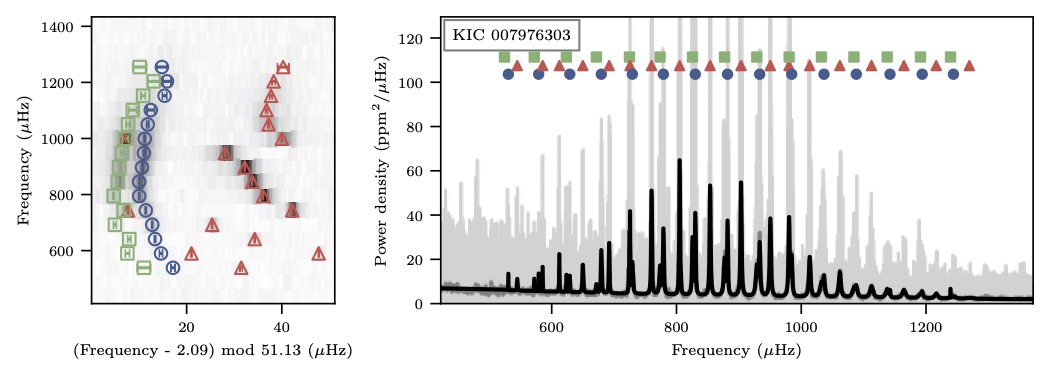} 
\includegraphics[width=0.94\linewidth]{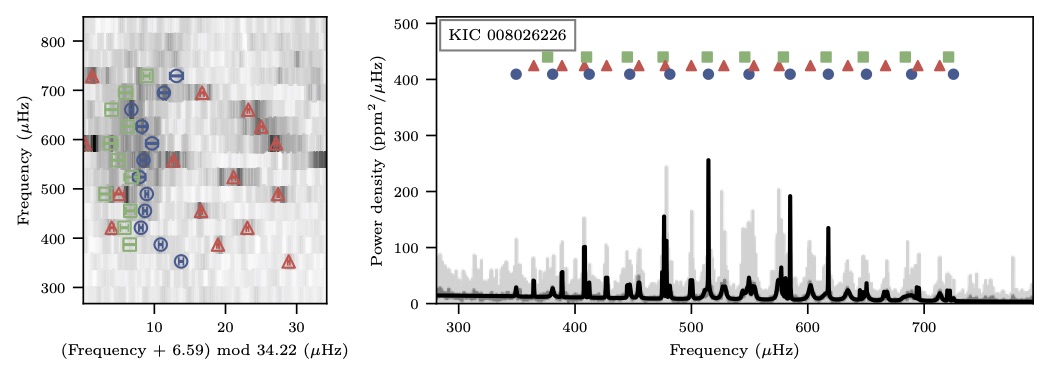} 
\includegraphics[width=0.94\linewidth]{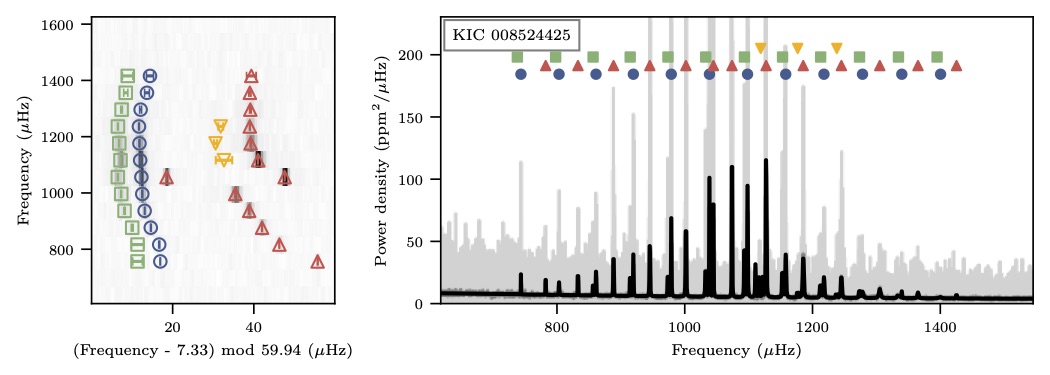} 
\includegraphics[width=0.94\linewidth]{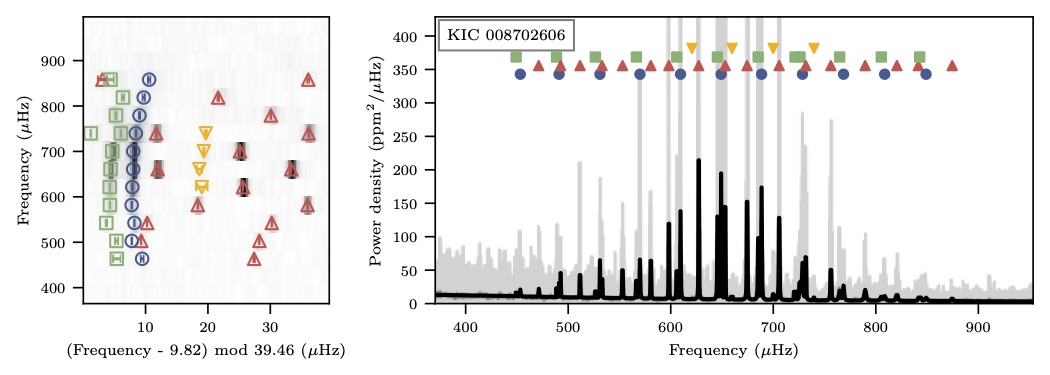} 
\end{figure*} 
\clearpage 
\begin{figure*} 
\includegraphics[width=0.94\linewidth]{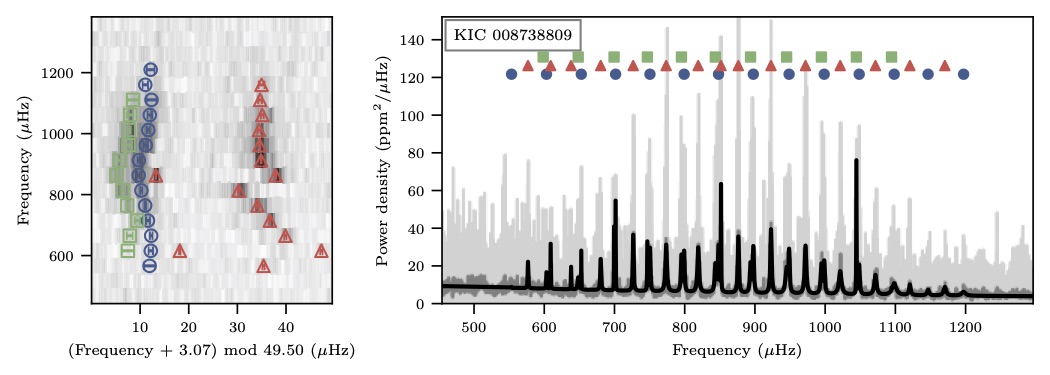} 
\includegraphics[width=0.94\linewidth]{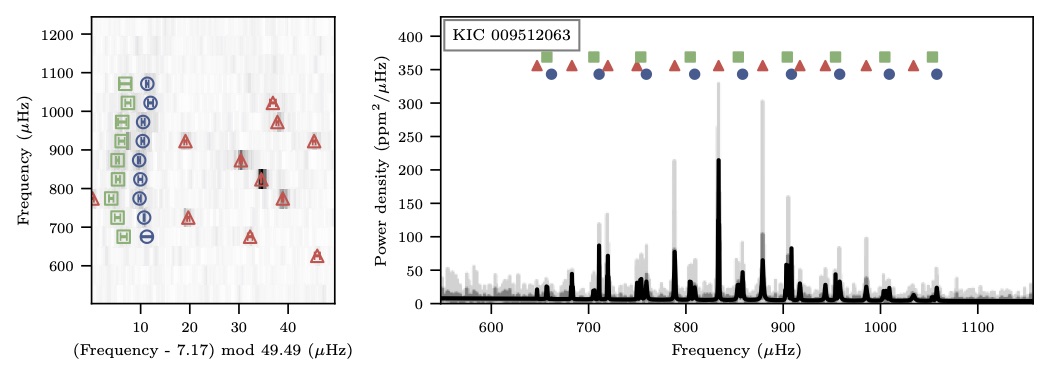} 
\includegraphics[width=0.94\linewidth]{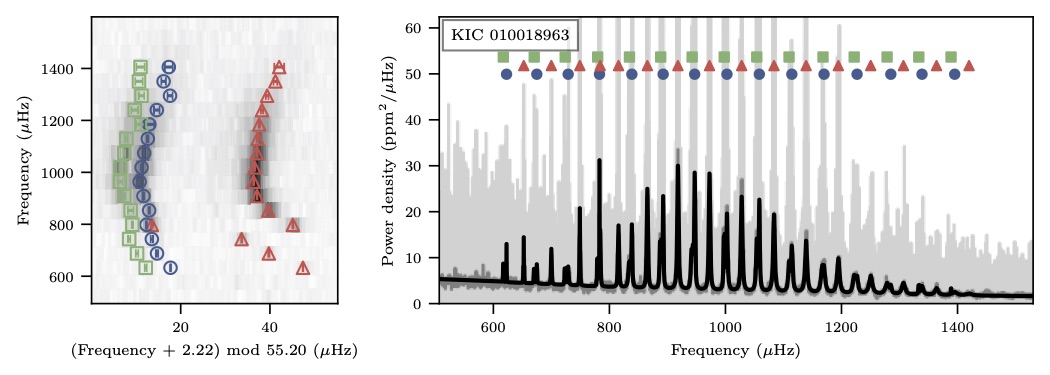} 
\includegraphics[width=0.94\linewidth]{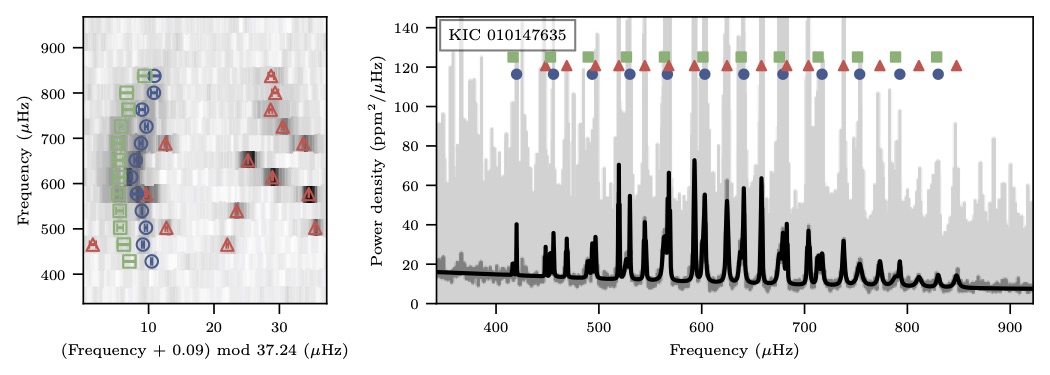} 
\end{figure*} 
\clearpage 
\begin{figure*} 
\includegraphics[width=0.94\linewidth]{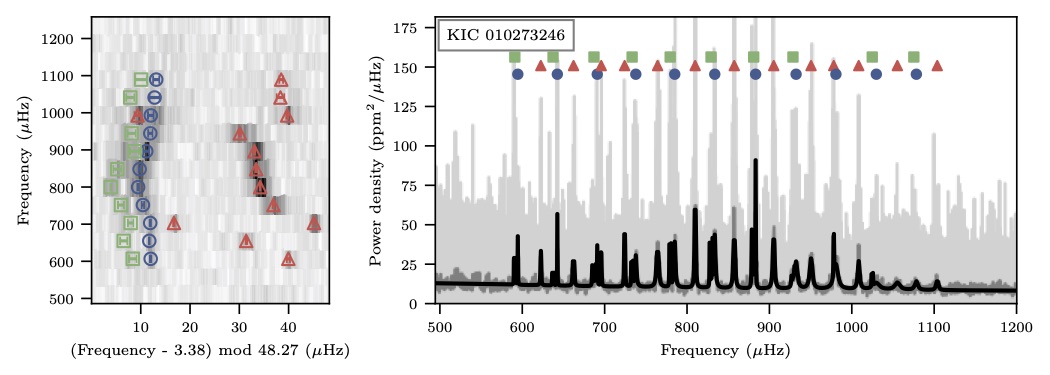} 
\includegraphics[width=0.94\linewidth]{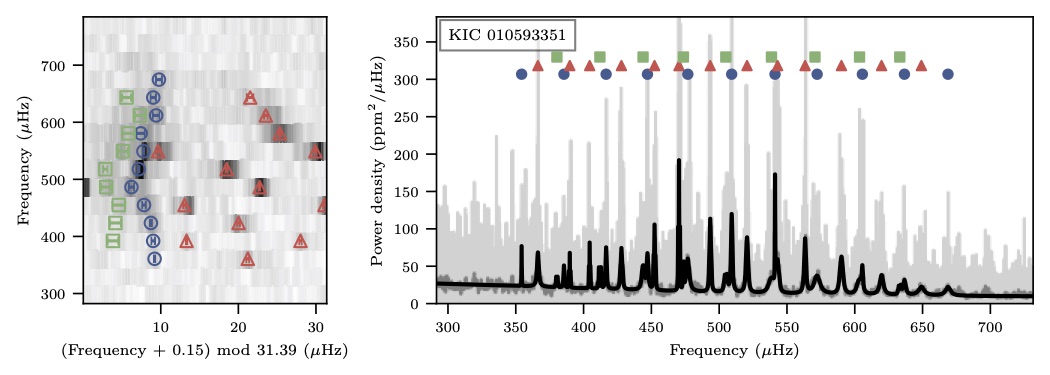} 
\includegraphics[width=0.94\linewidth]{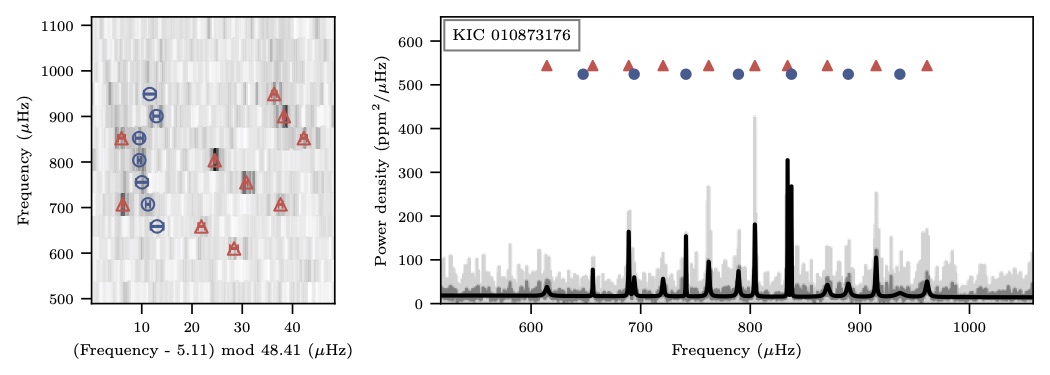} 
\includegraphics[width=0.94\linewidth]{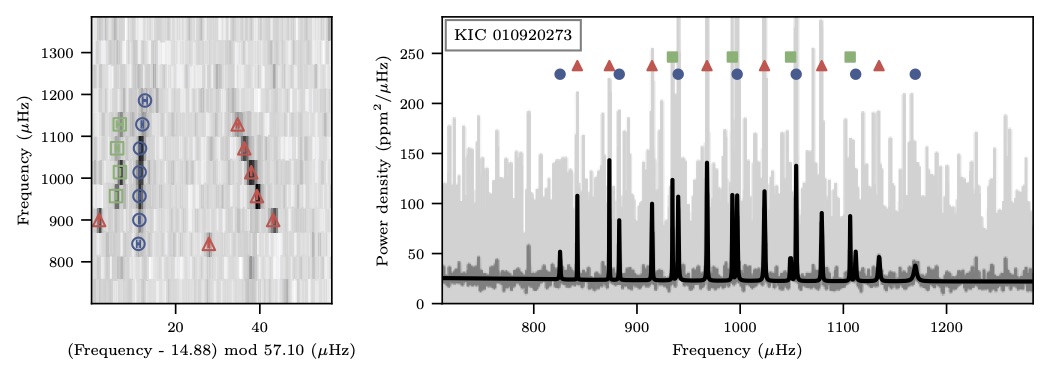} 
\end{figure*} 
\clearpage 
\begin{figure*} 
\includegraphics[width=0.94\linewidth]{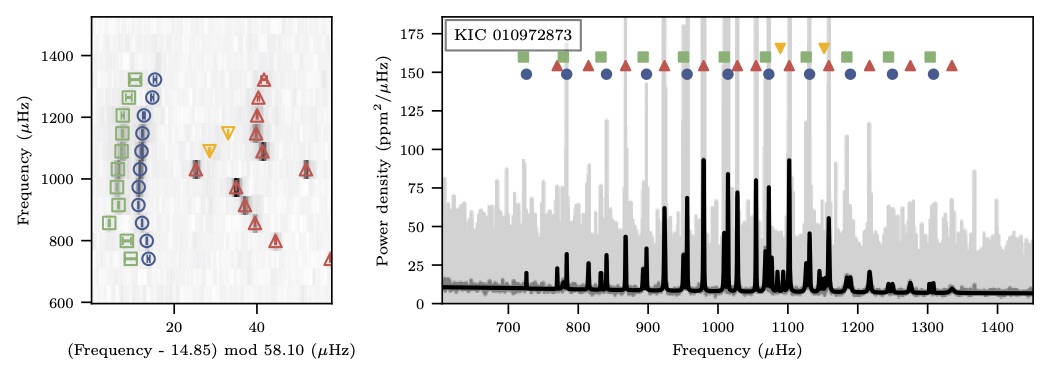} 
\includegraphics[width=0.94\linewidth]{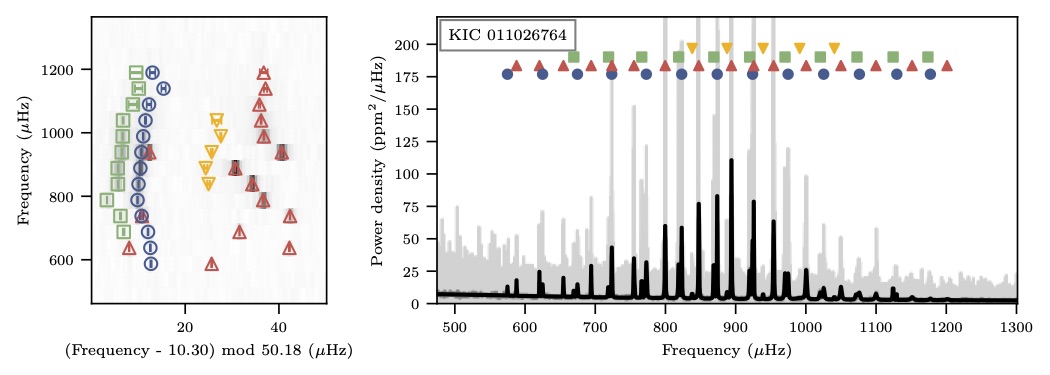} 
\includegraphics[width=0.94\linewidth]{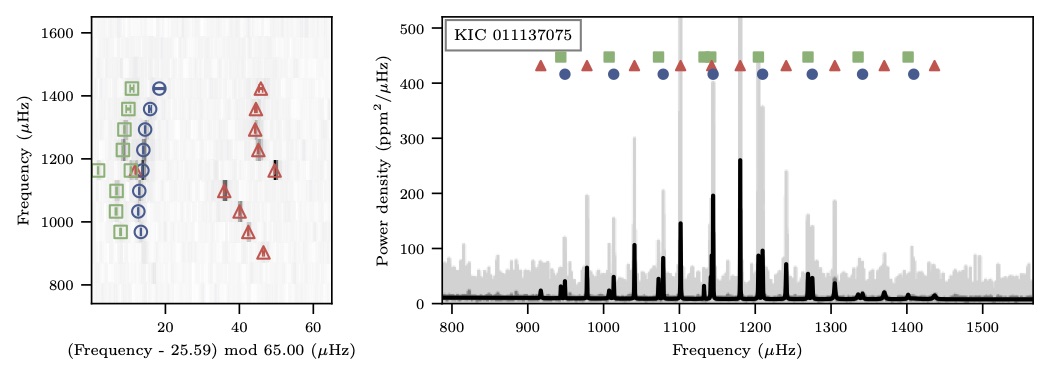} 
\includegraphics[width=0.94\linewidth]{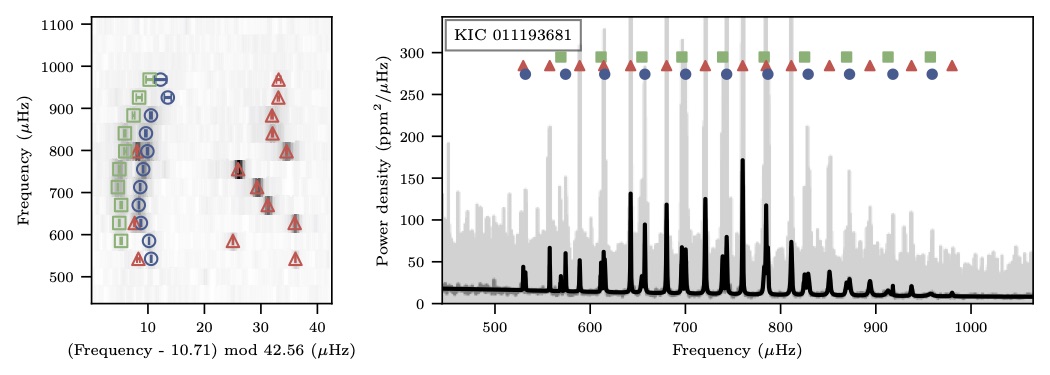} 
\end{figure*} 
\clearpage 
\begin{figure*} 
\includegraphics[width=0.94\linewidth]{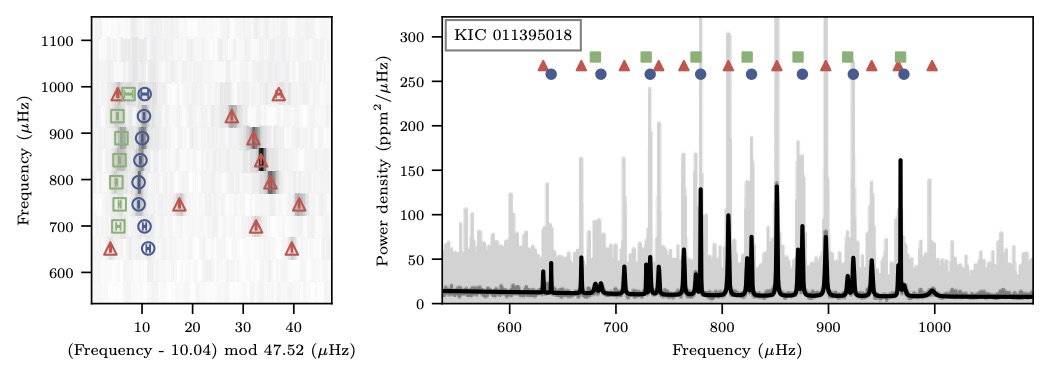} 
\includegraphics[width=0.94\linewidth]{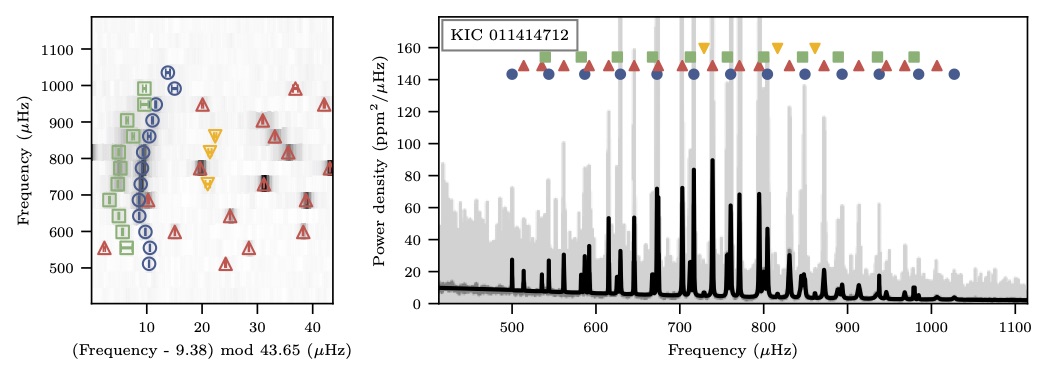} 
\includegraphics[width=0.94\linewidth]{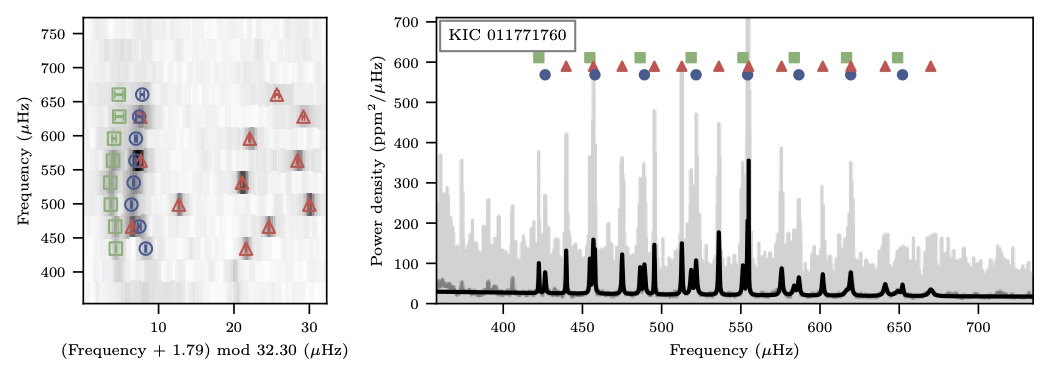} 
\includegraphics[width=0.94\linewidth]{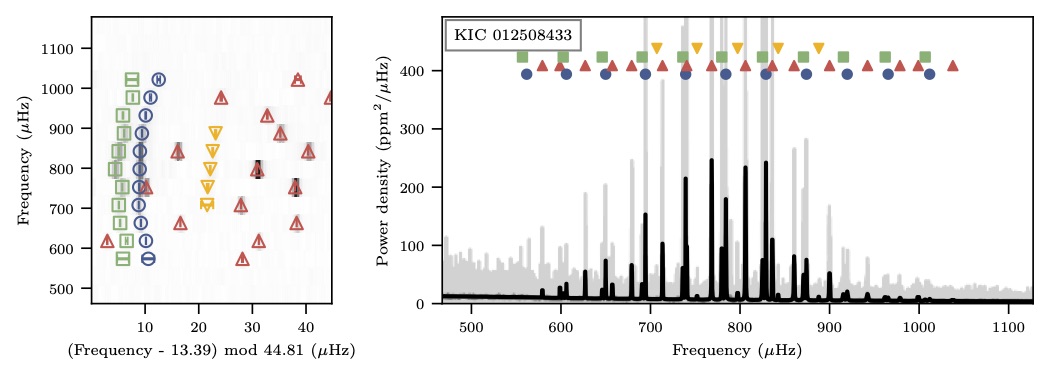} 
\end{figure*} 
\clearpage

\clearpage
\onecolumn
Table 2. Mode parameters obtained from the peakbagging. These are $m=0$ modes.

\begin{longtable}{rrrrrr}

\hline
      KIC & $l$ &     $\nu_{nl}$ ($\mu$Hz) & $\Gamma_{nl}$ ($\mu$Hz) &        $A_{nl}$ (ppm) &               $\ln K$ \\
\hline
\endfirsthead

\hline
      KIC & $l$ &     $\nu_{nl}$ ($\mu$Hz) & $\Gamma_{nl}$ ($\mu$Hz) &        $A_{nl}$ (ppm) &               $\ln K$ \\
\hline
\endhead

\hline \multicolumn{6}{r}{\textit{Continued on next page}} \\
\endfoot

\hline
\endlastfoot

 2991448 &   0 &   $889.34$ $\pm$ $0.312$ &     $0.79$ $\pm$ $0.96$ &   $4.04$ $\pm$ $1.08$ &     $2.16$ $\pm$ $0.75$ \\
  2991448 &   0 &   $949.46$ $\pm$ $1.111$ &     $8.02$ $\pm$ $2.37$ &   $9.16$ $\pm$ $1.13$ &     $3.61$ $\pm$ $0.98$ \\
  2991448 &   0 &  $1011.17$ $\pm$ $0.202$ &     $1.15$ $\pm$ $0.46$ &   $6.69$ $\pm$ $0.87$ &    $20.49$ $\pm$ $4.32$ \\
  2991448 &   0 &  $1072.78$ $\pm$ $0.099$ &     $0.76$ $\pm$ $0.24$ &  $10.45$ $\pm$ $1.06$ &    $39.42$ $\pm$ $2.54$ \\
  2991448 &   0 &  $1134.18$ $\pm$ $0.113$ &     $0.89$ $\pm$ $0.28$ &  $10.05$ $\pm$ $0.98$ &    $47.86$ $\pm$ $2.81$ \\
  2991448 &   0 &  $1195.89$ $\pm$ $0.547$ &     $3.29$ $\pm$ $1.72$ &   $9.45$ $\pm$ $1.68$ &     $5.40$ $\pm$ $2.13$ \\
  2991448 &   0 &  $1257.81$ $\pm$ $0.969$ &     $3.53$ $\pm$ $2.63$ &   $7.10$ $\pm$ $2.09$ &     $2.68$ $\pm$ $1.52$ \\
  2991448 &   0 &  $1319.52$ $\pm$ $1.294$ &     $3.00$ $\pm$ $3.56$ &   $3.60$ $\pm$ $2.33$ &     $0.18$ $\pm$ $1.20$ \\
\end{longtable}

\end{document}